\documentclass[twocolumn]{aastex6}

\usepackage{verbatim}
\usepackage{aas_macros}
\usepackage{hyperref}
\usepackage{natbib}
\usepackage{graphicx}
\usepackage{epsfig}
\usepackage{amsmath}
\usepackage[english]{babel}

\shorttitle{MONDian gravity in pressure supported systems}
\shortauthors{R. Durazo, X. Hernandez, B. Cervantes Sodi \& S. F. S\'anchez}

\begin{document}


\title{A universal velocity dispersion profile for pressure supported systems: evidence for MONDian gravity 
across 7 orders of magnitude in mass}


\author{R. Durazo$^{1}$, X. Hernandez$^{1}$, B. Cervantes Sodi$^{2}$ and S. F. S\'anchez$^{1}$}
\affil{$^{1}$Instituto de Astronom\'{\i}a, Universidad Nacional Aut\'{o}noma de M\'{e}xico, Apartado Postal 70--264 C.P. 
04510 M\'exico D.F. M\'exico.\\
$^{2}$Instituto de Radioastronom\'{\i}a y Astrof\'{\i}sica, Universidad Nacional Autonoma de M\'exico, Campus Morelia, 58090
Morelia, Michoac\'an, M\'exico.\\}



\begin{abstract}
For any MONDian extended theory of gravity where the rotation curves of spiral galaxies are explained through a change 
in physics rather than the hypothesis of dark matter, a generic dynamical behavior is expected for pressure 
supported systems: an outer flattening of the velocity dispersion profile occurring at a characteristic radius, where 
both the amplitude of this flat velocity dispersion and the radius at which it appears are predicted to show distinct 
scalings with the total mass of the system. By carefully analyzing the dynamics of globular clusters and elliptical galaxies,
we are able to significantly extend the astronomical diversity of objects in which MONDian gravity
has been tested, from spiral galaxies, to the much larger mass range covered by pressure supported systems.
We show that a universal projected velocity dispersion profile accurately describes various classes of pressure supported 
systems, and further, that the expectations of extended gravity are met, across seven orders of magnitude in mass. 
These observed scalings are not expected under dark matter cosmology, and would require particular explanations tuned at 
the scales of each distinct astrophysical system.
\end{abstract}


\section{Introduction}

Within the context of modified gravity theories constructed to yield galactic dynamics in the absence of
dark matter, one of the most convincing supporting arguments comes from the success of MOND in reproducing 
rotation curves of spiral galaxies (Milgrom 1983; Swaters et al. 2010; McGaugh 2012). Both the 
flatness of the observed spiral rotation curves and the scaling of the asymptotic rotation with the fourth 
root of the total baryonic content, i. e. the Tully-Fisher relation, can be accurately reproduced across spiral 
galaxy sizes and types, as well as the overall shapes and correlations between observed features in the baryonic 
distribution and small kinematic irregularities (Sanders \& McGaugh 2002; Famaey \& McGaugh 2012; 
Rodrigues 2014; Lelli et al. 2016). However, such successful tests are restricted to the range of mass scales
and morphology of spiral galaxies.

To test the validity (or lack thereof) of MONDian gravity schemes, it is desirable to extend the diversity of
astrophysical systems over which these ideas can be probed. To do this, we give the expected kinematic profiles
for two different types of pressure supported systems, astrophysical objects with baryonic masses spanning from $10^{4} M_{\sun}$
for globular clusters, to $10^{11} M_{\sun}$ for elliptical galaxies. 

By the term MONDian gravity we shall refer to any extended gravity theory where at $a>a_{0}$ scales Newtonian 
gravity is recovered, while the $a<a_{0}$ regime reproduces MOND dynamics, at the $v<<c$ limit, where $a_{0}$ is 
Milgrom's acceleration of $1.2 \times 10^{-10} m/s^{2}$ (Bekenstein 2004; Capozziello \& De Laurentis 2011;  
Mendoza et al. 2013; Verlinde 2016).  Generically for such theories, beyond a radius 
given by $R_{M}= (GM /a_{0})^{1/2}$, centrifugal equilibrium velocities will become flat at a level of 
$V=(GM a_{0})^{1/4}$ (Milgrom 1983). Similarly, for pressure supported systems, beyond around $R_{M}$, velocity dispersion
profiles will stop falling along Newtonian expectations and flatten out at a level of $\sigma_{\infty}=V/3^{1/2}$,
with $M$ the total baryonic mass of a rapidly converging astrophysical system (Milgrom 1984; Hernandez \&
Jimenez 2012).

Recently, McGaugh et al. (2016) and Lelli et al. (2016) used the SPARC rotation curve and photometry databese, to
show that the baryonic mass distribution alone directly determines the resulting acceleration, along Newtonian
expectations in the high acceleration regime, and along MONDian ones when $a<a_{0}$.
A similar study reaching conclusions also in support of MONDian dynamics, but
using structural, dynamical and mass determinations for a range of elliptical
galaxies is that of Dabringhausen et al. (2016).
Baryonic mass determination
in the above studies however, are subject to numerous  assumptions, as well as uncertainties and
systematics, which vary for differing classes of astrophysical systems. Hard to determine gas fractions, star
formation histories, the presence of multiple and complex stellar populations and unknown stellar mass functions
imply mass to light ratios and their radial variations are subject to large empirical errors.

Here we take a complementary approach, using only kinematical data and avoiding any dynamical physical assumptions or
mass estimates in deriving the parameters we study, concentrating merely on the shape of observed projected velocity
dispersion profiles. The expected theoretical scalings of both $R_{M}$ and $\sigma_{\infty}$ with $M$ imply the mass can be
eliminated to obtain $R_{M} \propto \sigma_{\infty}^{2}$, a relation between the parameters of the directly observable
velocity dispersion profiles which can now be directly tested.

By analysing velocity dispersion profiles from Galactic globular clusters recently observed by Scarpa et al. (2007a; 2007b), 
Scarpa \& Falomo (2010) and Scarpa et al. (2011), henceforth the Scarpa et al. group, and Lane et al. (2009), Lane et al. 
(2010a; 2007b) and Lane et al. (2011), henceforth the Lane et al. group, and elliptical galaxies having high quality 2D
kinematics by S\'anchez et al. (2016b), using data from the CALIFA project, (S\'anchez et al. 2012;
Walcher et al. 2014; S\'anchez et al. 2016a),  we show that a universal projected velocity dispersion profile 
$\sigma(R)= \sigma_{0}e^{-(R/R_{\sigma})^{2}} +\sigma_{\infty}$ adequately serves to model the two distinct classes of 
astrophysical systems. Further, a scaling of $R_{M} \propto \sigma_{\infty}^{n}$ ensues, with $n=1.89 \pm 0.32$, compatible 
with the generic expectations of MONDian gravity models of $n=2$.

In section (2) we present the development of simple first order predictions for the large scale velocity
dispersion expected under MONDian gravity, and the corresponding scalings between the parameters
describing such velocity dispersion profiles for pressure supported systems. Section (3) includes a description
of the data used for sets of velocity dispersion profiles for globular clusters and elliptical galaxies, together
with the fitting procedure and parameters obtained for the universal velocity dispersion profile
proposed. In section (4) we show how the two sets of systems treated, spanning seven orders of magnitude 
in mass, follow the proposed universal velocity dispersion profile, and show also how the flattening
radii follow an overall scaling consistent with being proportional to the square of the asymptotic velocity dispersion, 
as predicted generically under MONDian gravity. Section (5) summarizes our conclusions.

\section{MONDian Theoretical expectations}

Thinking of modified gravity in terms of a Newtonian force law implies a change of regime from the Newtonian
expression of $F_{N}=GM/r^{2}$ to $F_{M}=(GM a_{0})^{1/2}/r$, occurring at scales of $R_{M}=(G M /a_{0})^{1/2}$,
if one wants to model observed galactic dynamics in the absence of any dark matter (Milgrom 1983). 
The details of the transition are not clear, beyond the requirements from consistency with 
solar system dynamics (Mendoza et al. 2011), Milky Way rotation curve comparisons (Famaey \& Binney 2005)
and Galactic globular cluster modelling (Hernandez et al. 2012), for the transition to be fairly abrupt. 
Under such schemes, centrifugal equilibrium velocities will tend to the Tully-Fisher value of:

\begin{equation}
V=\left( G M a_{0}\right)^{1/4},
\end{equation}

\noindent for test particles orbiting a total baryonic mass M. Under the MONDian regime, the relation between
centrifugal equilibrium velocities and isotropic velocity dispersion will be only slightly modified with respect to 
the Newtonian case, to yield:

\begin{equation}
 \sigma_{\infty} = V/ \sqrt{3},
\end{equation}

\noindent e.g.  Milgrom (1984), Hernandez \& Jimenez (2012). Thus, we can write the isotropic Maxwellian velocity
dispersion of a pressure supported system in the MONDian regime as:

\begin{equation}
\sigma_{\infty}^{2}= \frac{1}{3} \left( G M a_{0}\right)^{1/2}.
\end{equation}

In order to extend the range over which MONDian dynamics can be tested 
to the much larger ones over which pressure supported systems appear,  would require observations of
$\sigma_{\infty}$ and independent inferences of the total baryonic mass of various systems, to directly test the validity
of Equation (3). Any such undertaking will be hindered by the large theoretical uncertainties in total (and radially
changing) mass to light ratios, uncertain gas fractions, and details of the relevant IMFs and star formation histories,
which to make matters worse, vary significantly from the ``simple'' cases of globular clusters to the more complex
histories of elliptical galaxies. One can circumvent such uncertainties and obtain a simpler empirical test by eliminating
$M$ from the above equation in favor of $R_{M}$ to yield: 

\begin{equation}
R_{M}=\frac{3 \sigma_{\infty}^{2}}{a_{0}},
\end{equation}

\noindent which in astrophysical units reads:

\begin{equation}
\left( \frac{R_{M}}{pc}\right) = 0.81 \left( \frac{\sigma_{\infty}}{km/s} \right)^{2}.
\end{equation}

The above expression now relates only features of the kinematic velocity dispersion profile of an astronomical
system, as $\sigma_{\infty}$ will be the large scale asymptotic projected velocity dispersion profile, and 
$R_{M}$ will be the radial scale at which the velocity dispersion profile becomes flat. The above will only
make any sense if observed velocity dispersion profiles do indeed show such morphology, for a
decline in an inner Newtonian region, and a transition to a flat velocity dispersion regime on crossing the
$R_{M}$ threshold.

We are inspired to attempt such a comparison across the range of scales covered by
pressure supported systems by recent results coming from the observed velocity dispersion profiles of
Galactic globular clusters (Scarpa et al. 2010; Scarpa \& Falomo 2011) showing precisely such a morphology, 
and recently shown to comply with Tully-Fisher MONDian expectations (Hernandez et al. 2013a).
Further, Desmond (2017) recently demonstrated through an extensive sample of rotation
curve observations and baryonic mass inferences across sizes and galactic types, that the inferred mass discrepancy
acceleration relation is consistent with MONDian expectations.

In the following section we show a number of velocity dispersion profiles having precisely
the morphology described above, for Galactic globular clusters and elliptical galaxies,
which in Section 4 we compare to the expectations of Equation 5.

Notice also that multiplying Equations (3) and (4), and substituting $M$ on
the right hand side for $\sigma_{\infty}$, 
the resulting combination of parameters $R_{M} \sigma_{\infty}^2 =GM/3$
is highly reminiscent of the corresponding Newtonian expression, changing $R_{M}$ for the half mass radius,
and $\sigma_{\infty}$ for the central velocity dispersion of a purely Newtonian system (Cappellari et al. 2006; Wolf
et al. 2010).

In the above we have assumed that the total baryonic mass has essentially converged,
and on reaching the flat dispersion velocity region the potential of the
system in question can be treated as a point mass, in consistency with the
validity of Newton's theorems for spherically symmetric mass distributions also
under MONDian gravity (e.g. Mendoza et al. 2011), and that the dynamical
tracers from which $\sigma_{\infty }$ is calculated behave as test particles. Although this will never be 
strictly true, it will hold to a good approximation, given the very centrally concentrated mass profiles
of e.g. Galactic globular clusters and the inferred density profiles of close to the de Vaucouleurs surface brightness of
elliptical galaxies. Indeed, pressure supported systems have been accurately
modeled recently in the MONDian regime reproducing simultaneously the observed surface brightness profiles
and the measured velocity dispersion ones, e.g. Galactic globular clusters by Gentile et al. (2010),  
the giant elliptical galaxy NGC4649 by Jimenez et al. (2013) and tenuous 
galactic stellar halos by Hernandez et al. (2013b), all in absence of any otherwise dominant dark matter component.
More recently, Chae \& Gong (2015) studied the predicted distribution of velocity dispersion profiles (VDP) for elliptical
galaxies and found that their MONDian models could reproduce the observed distribution of VDP slopes requiring only
the observed stellar mass, without any dark matter, and Tian \& Ko (2016) concluded that the dynamics of seven
elliptical galaxies, as traced by stellar and planetary nebulae, could be well explained by MOND.
The issue remains controversial, as some studies have found dark matter halos to be a better fit
than MOND in elliptical galaxies (Richtler et al. 2008; Samurovi{\'c} 2014, 2016).

\begin{figure*}
\gridline{\fig{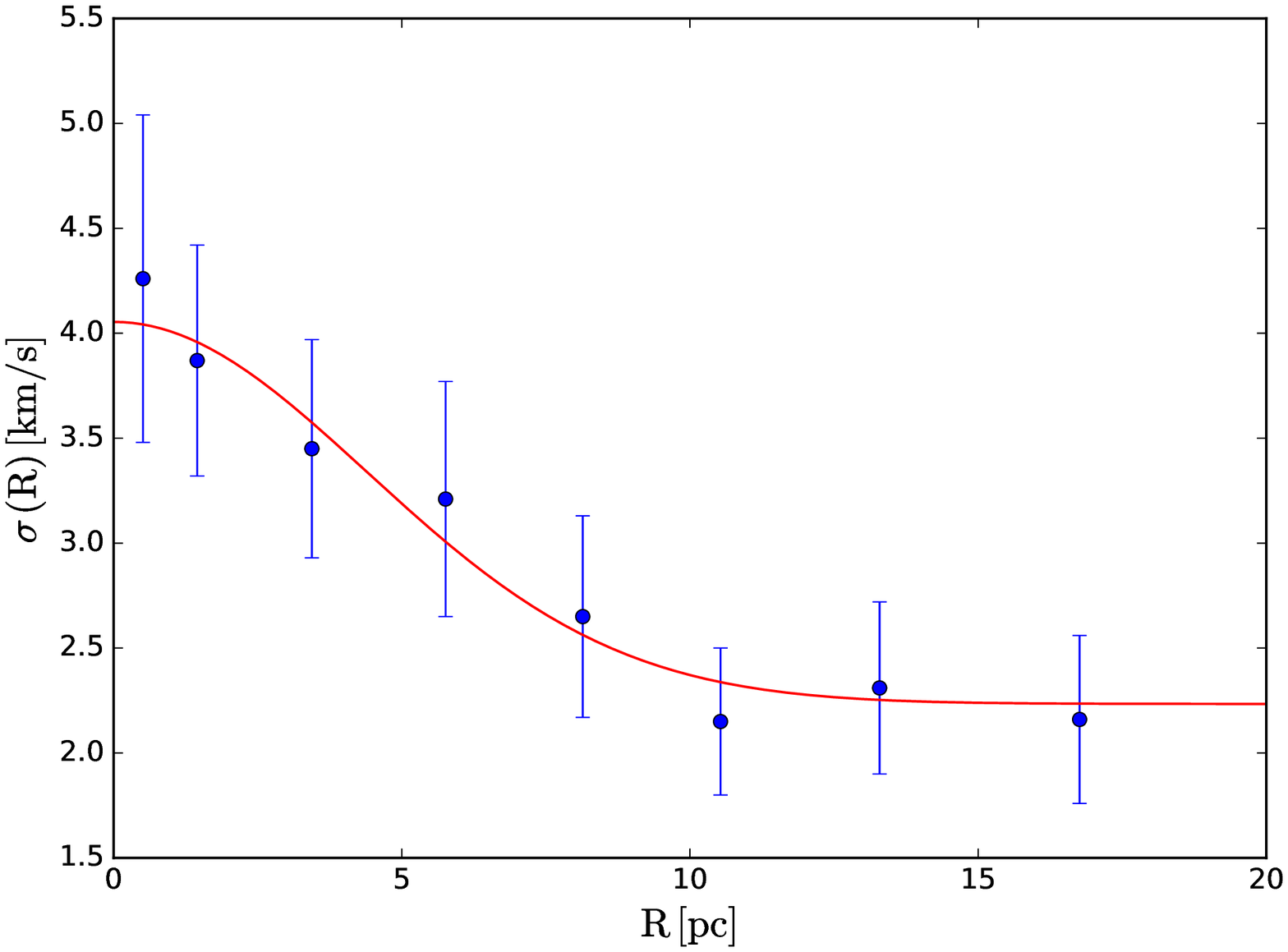}{0.3\textwidth}{NCG 7099}
          \fig{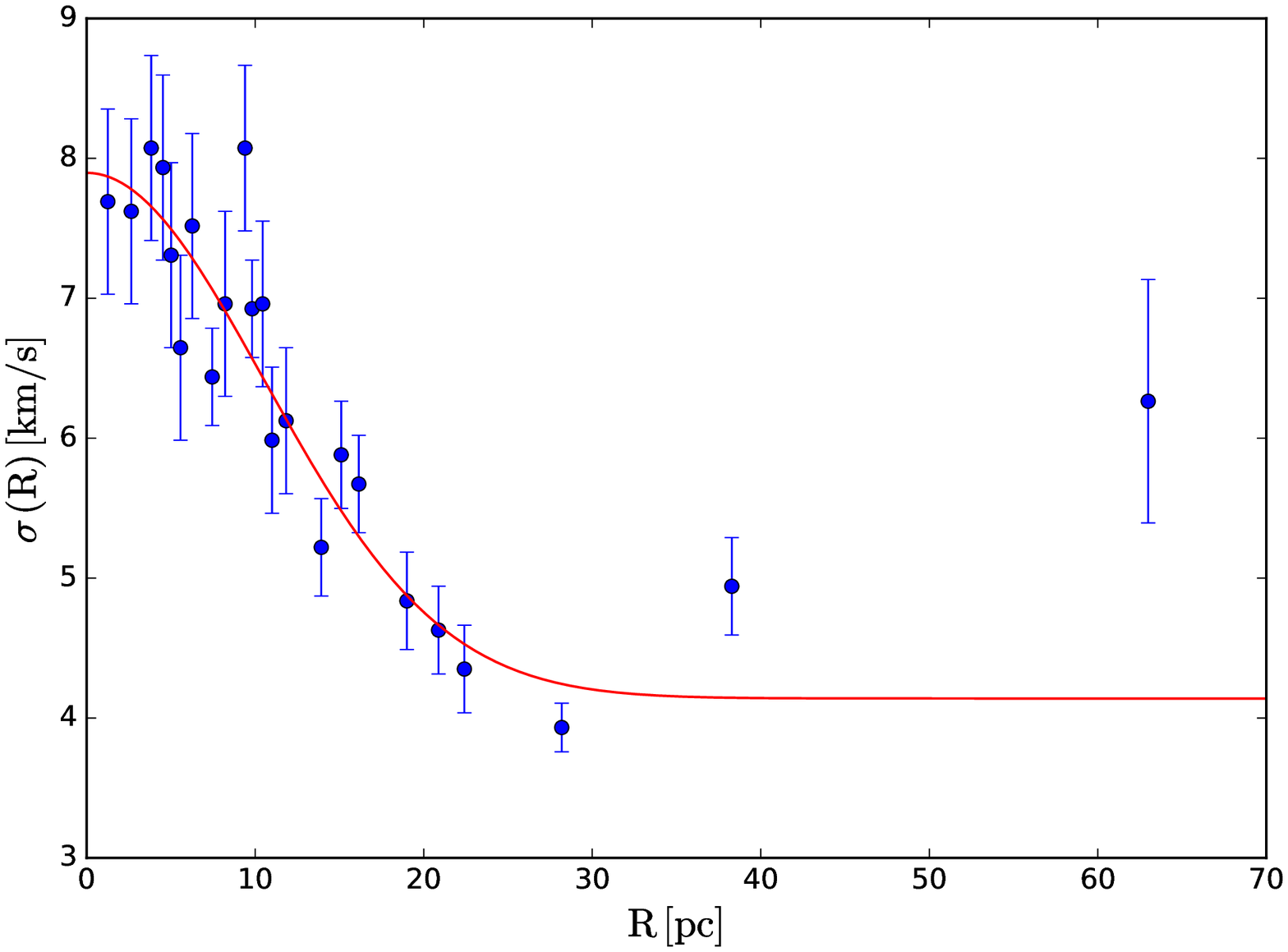}{0.3\textwidth}{NGC 0104}
          \fig{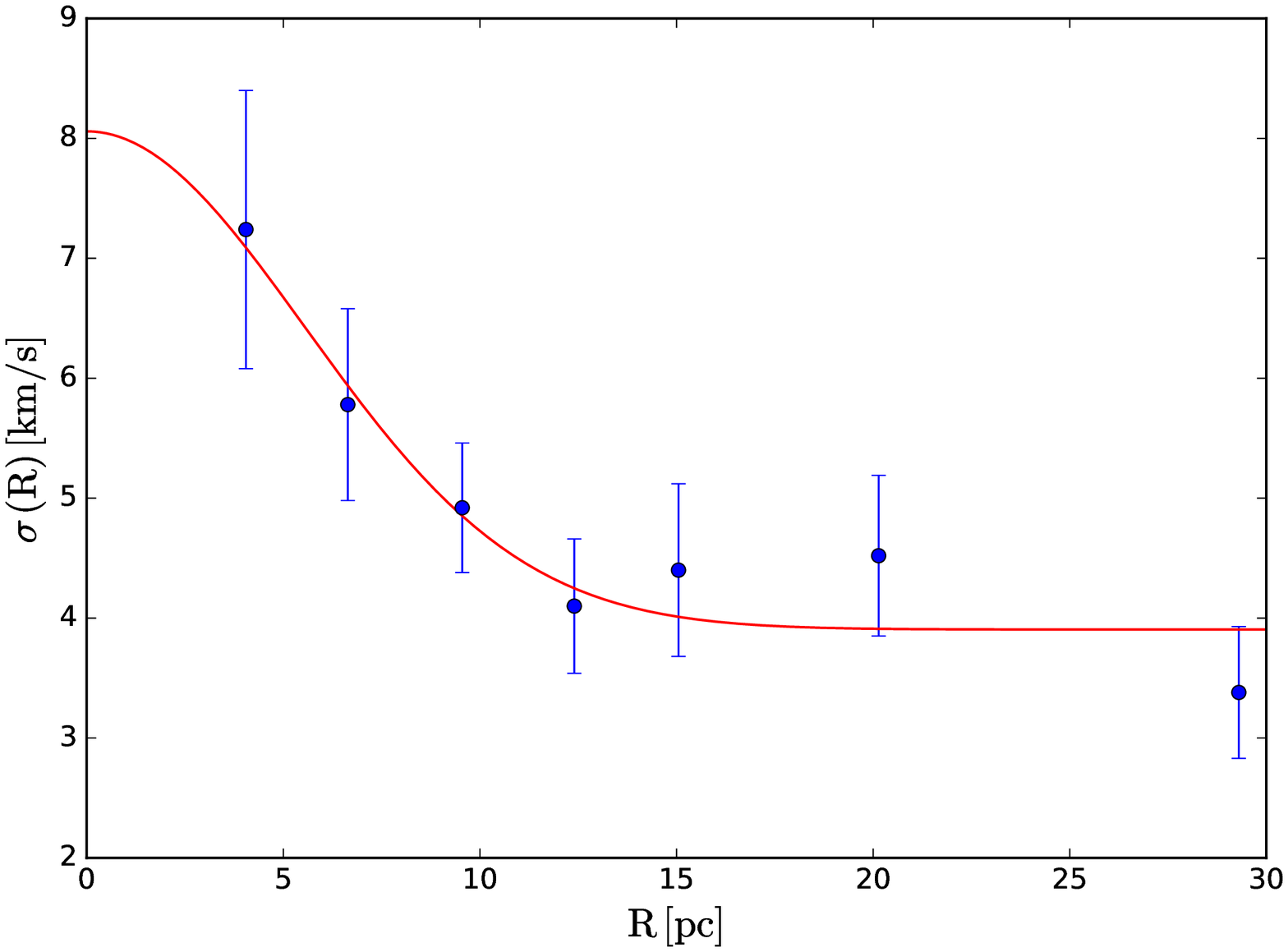}{0.3\textwidth}{NGC 1851}
          }
\gridline{\fig{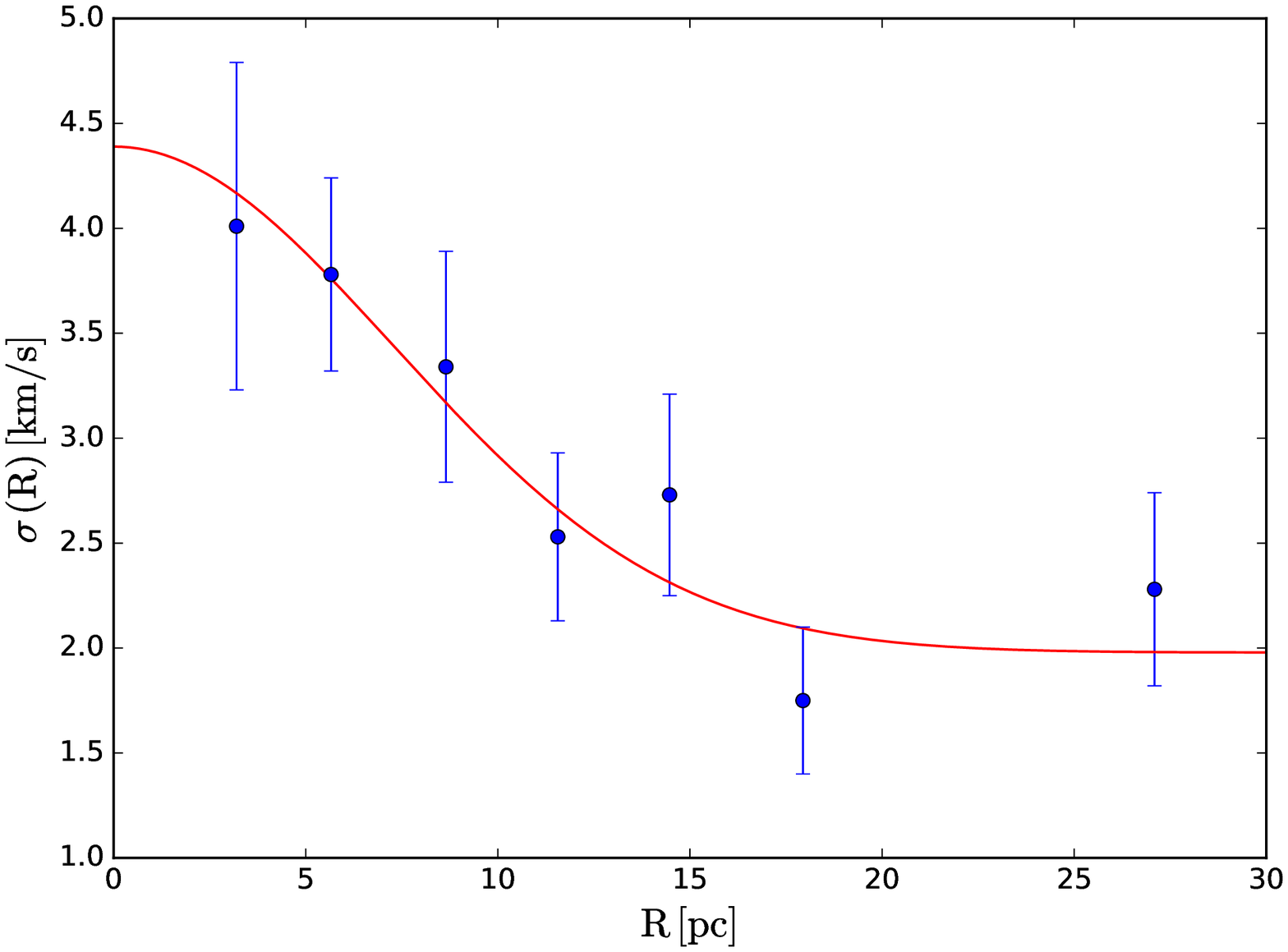}{0.3\textwidth}{NCG 1904}
          \fig{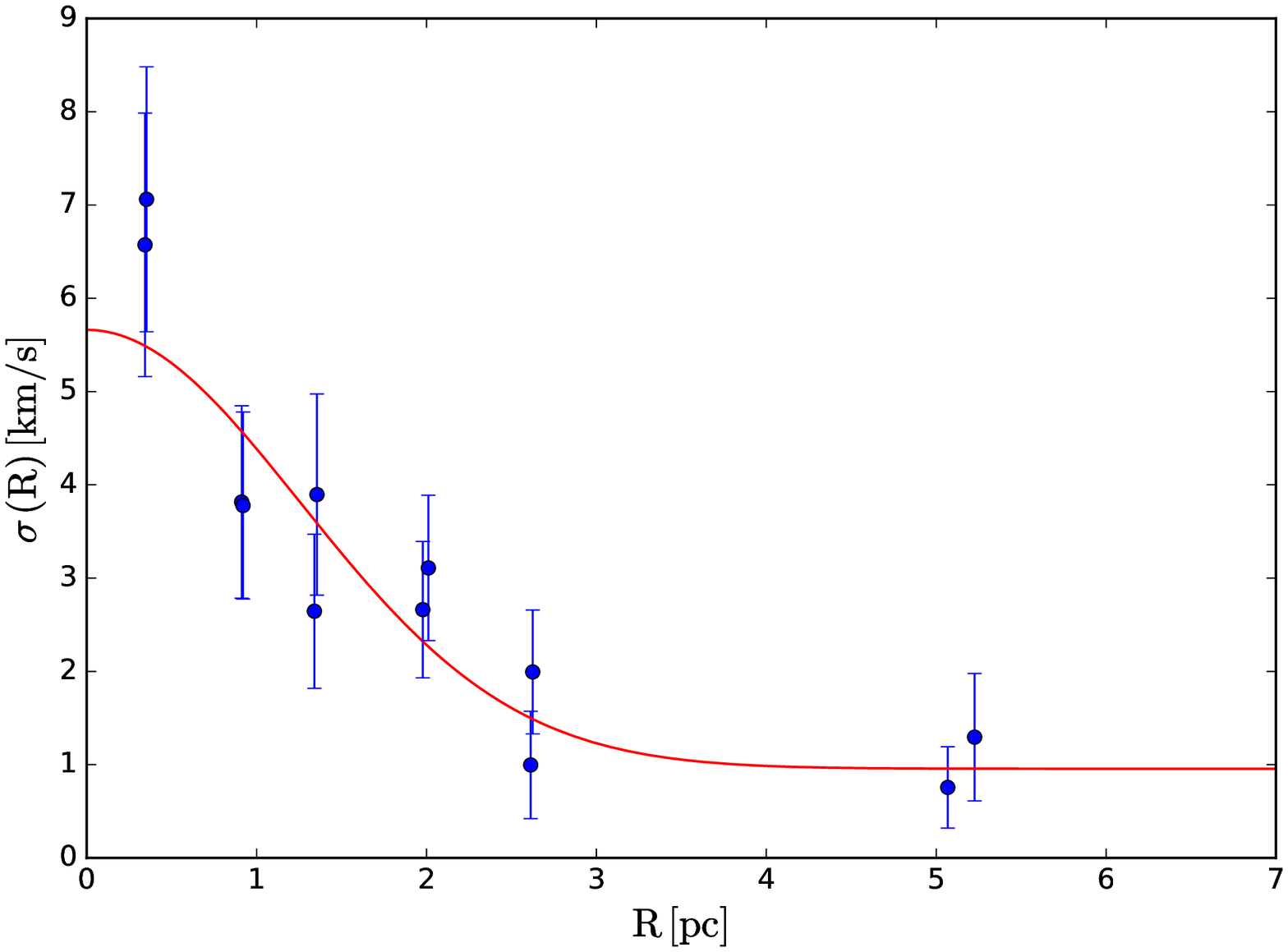}{0.3\textwidth}{NGC 2419}
          \fig{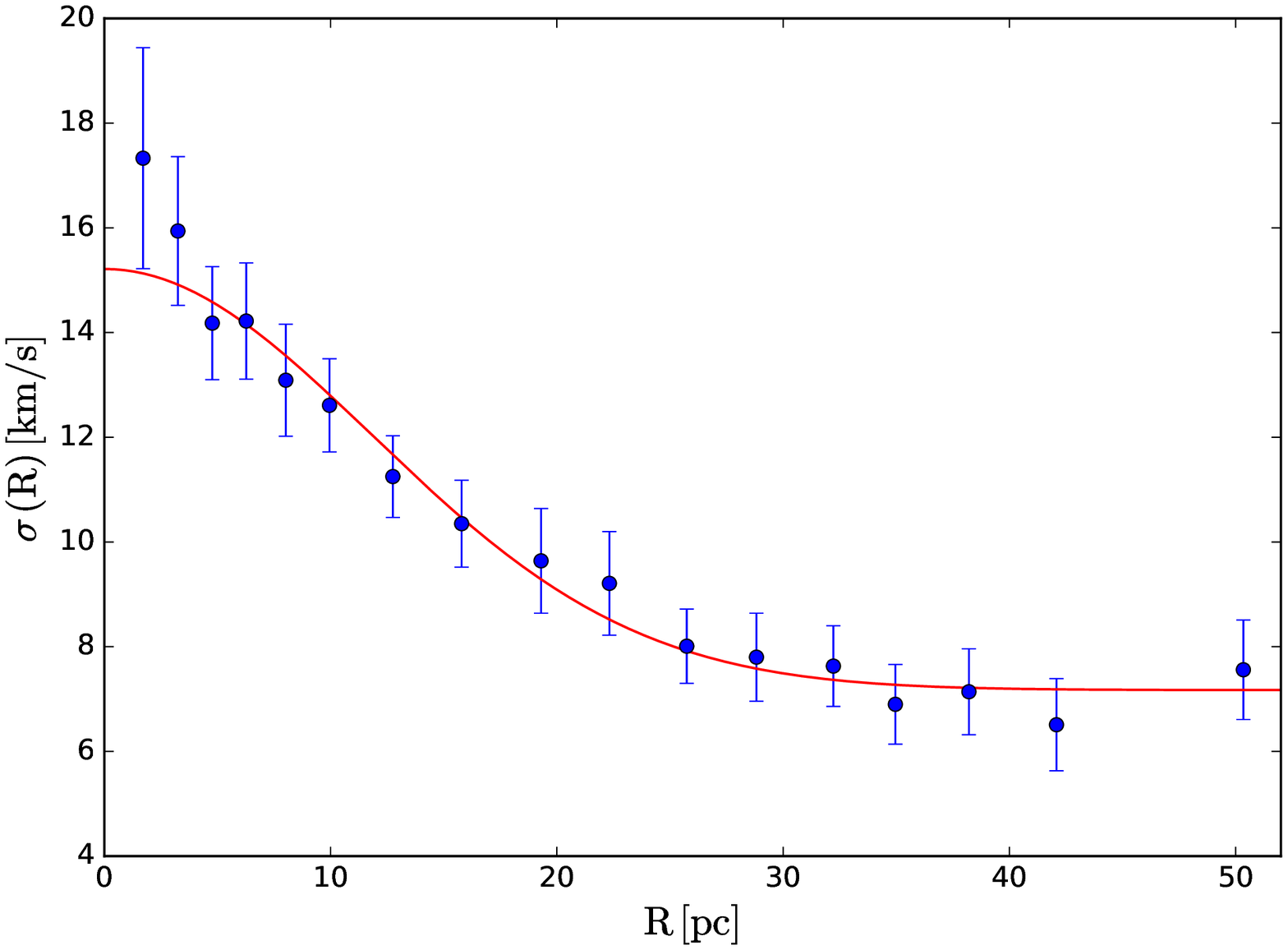}{0.3\textwidth}{NGC 5139}
          }
\gridline{\fig{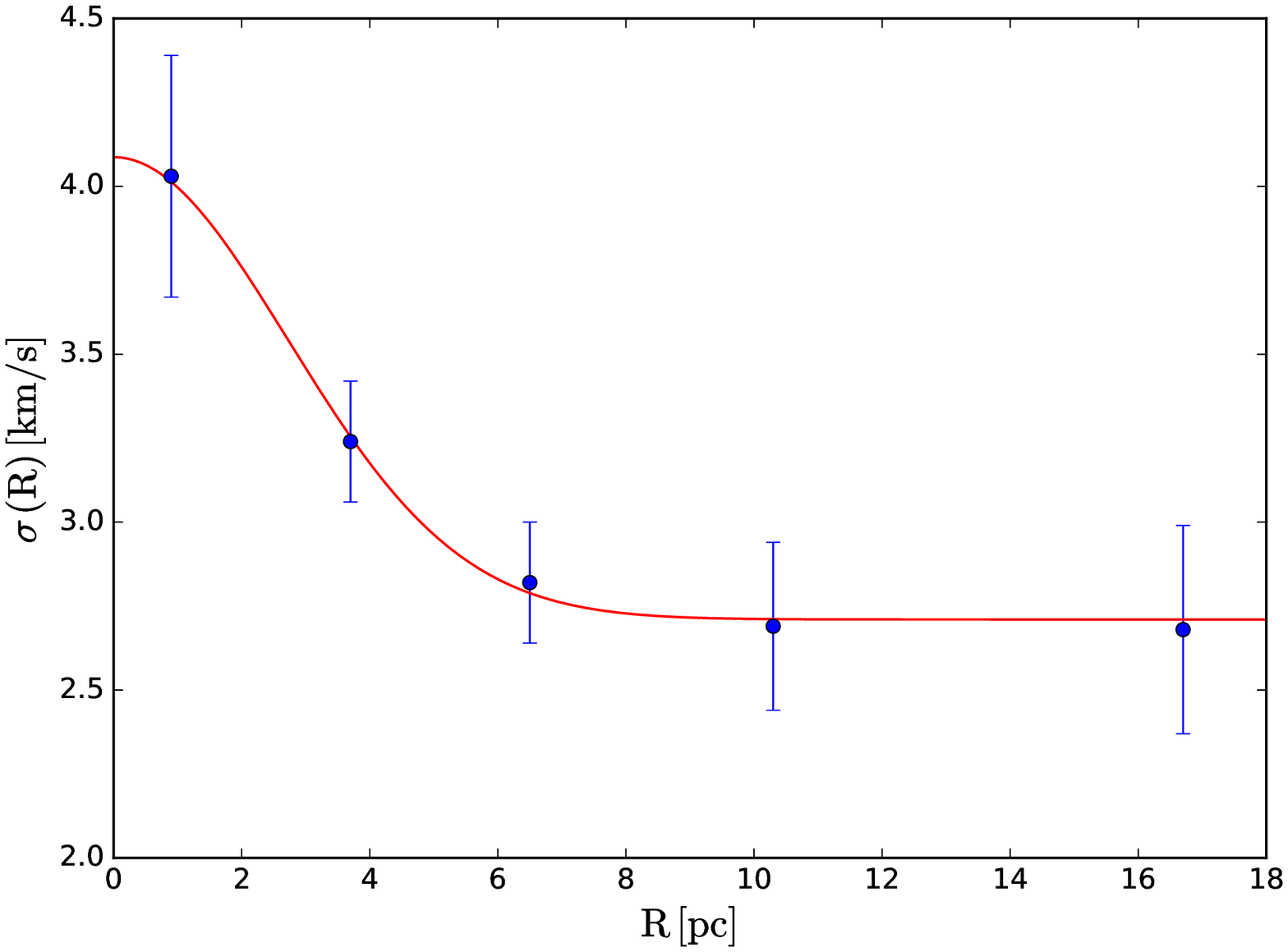}{0.3\textwidth}{NCG 6171}
          \fig{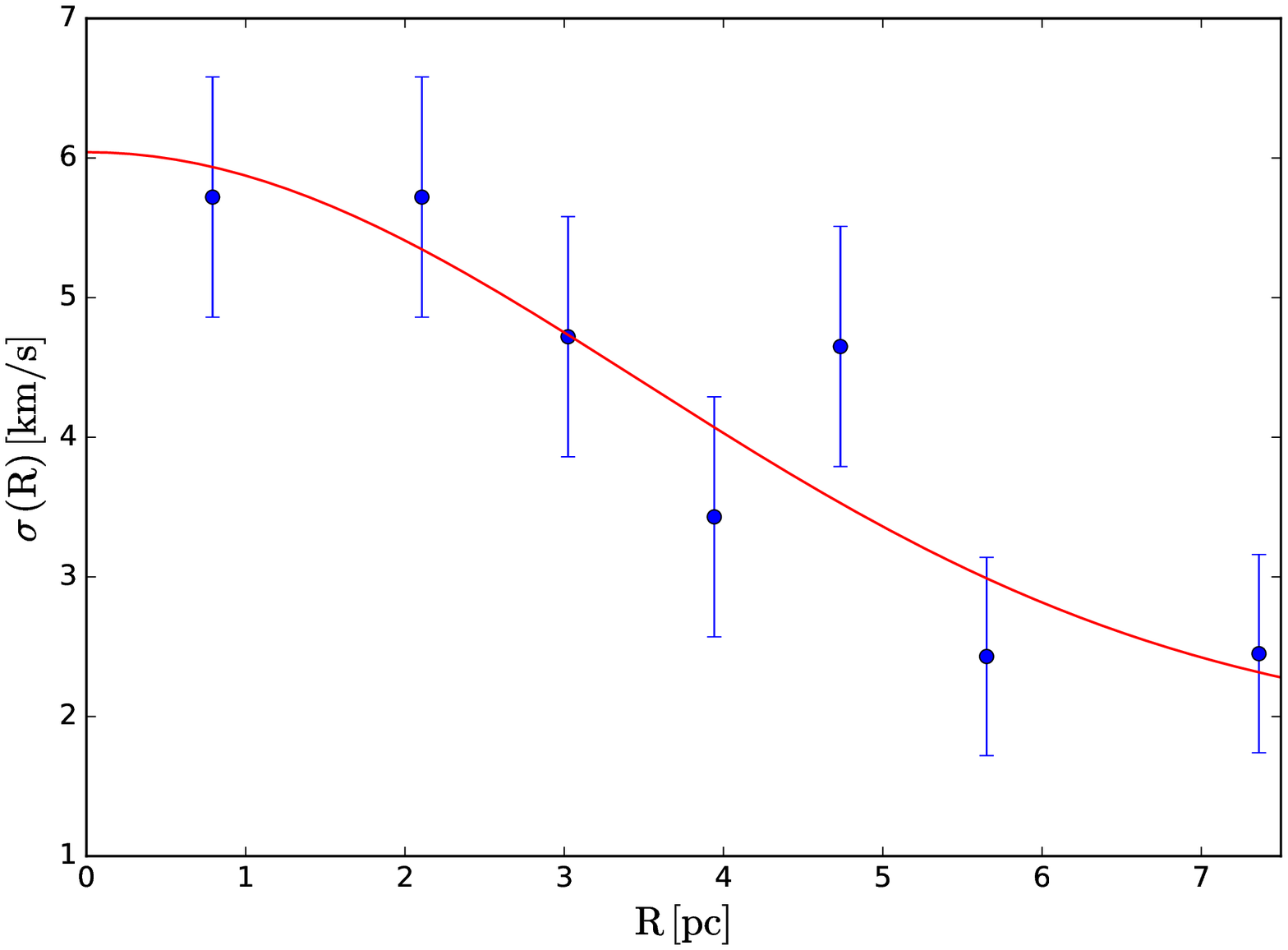}{0.3\textwidth}{NGC 6218}
          \fig{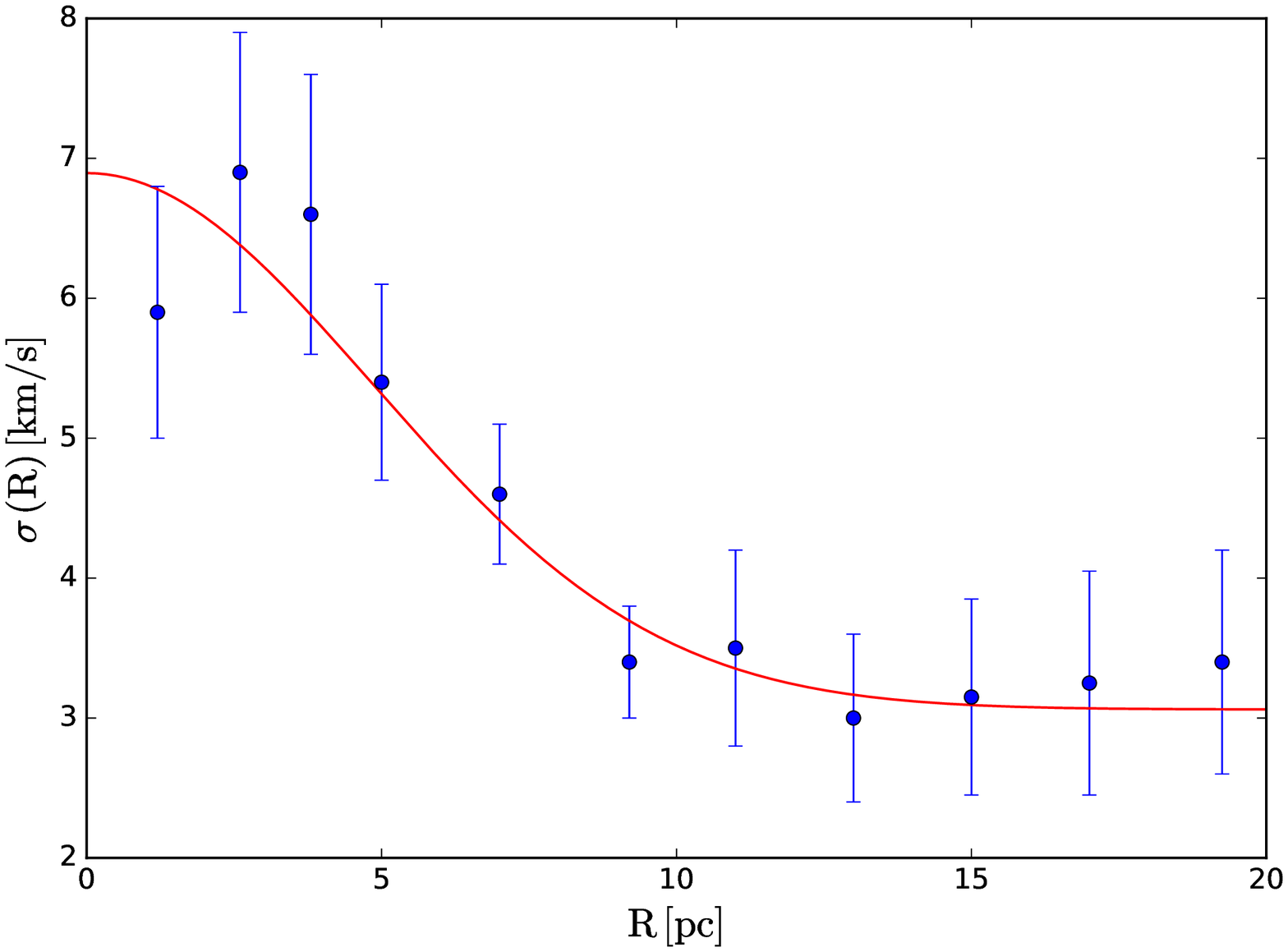}{0.3\textwidth}{NGC 6341}
          }
\gridline{\fig{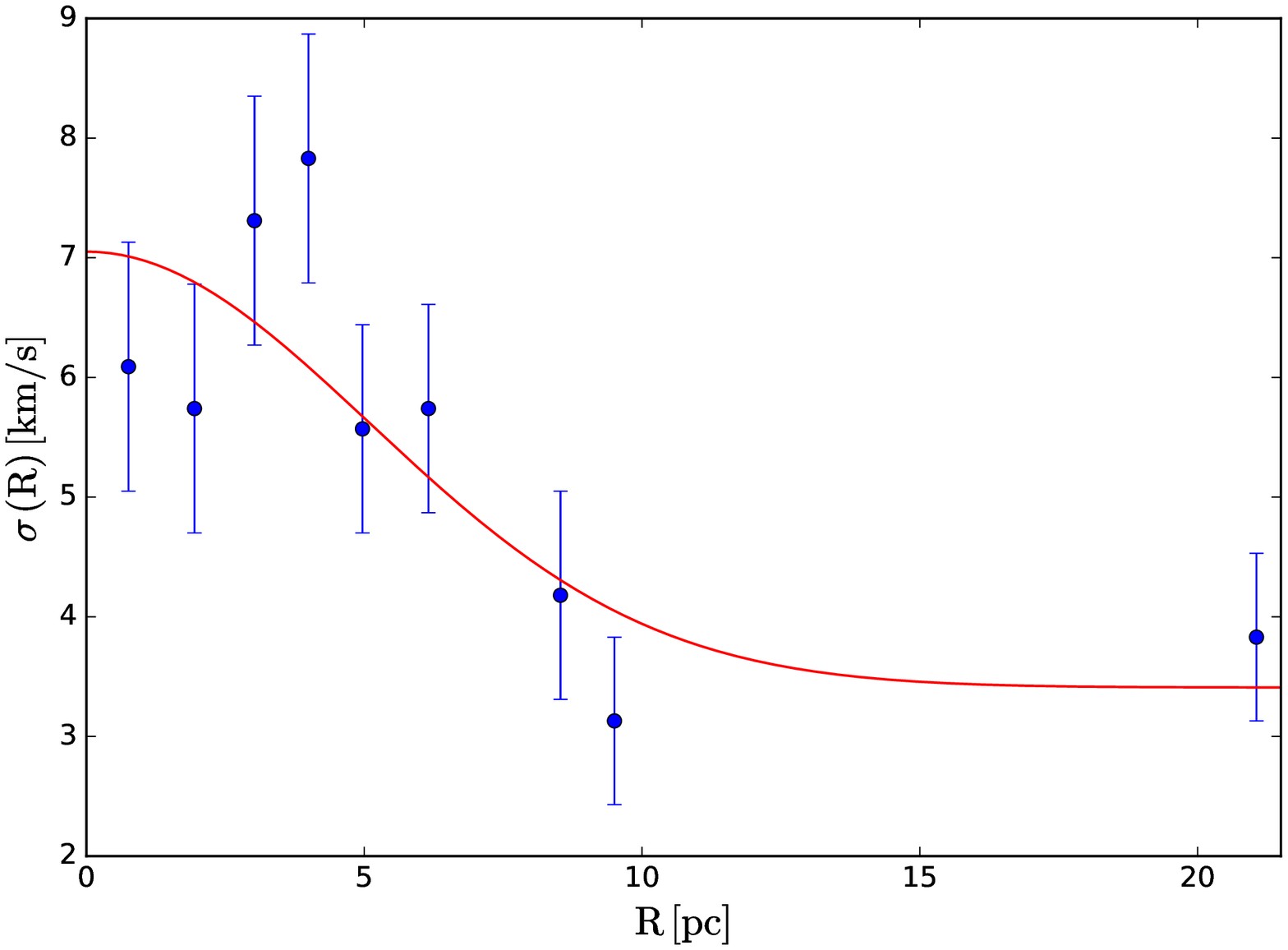}{0.3\textwidth}{NCG 6656}
          \fig{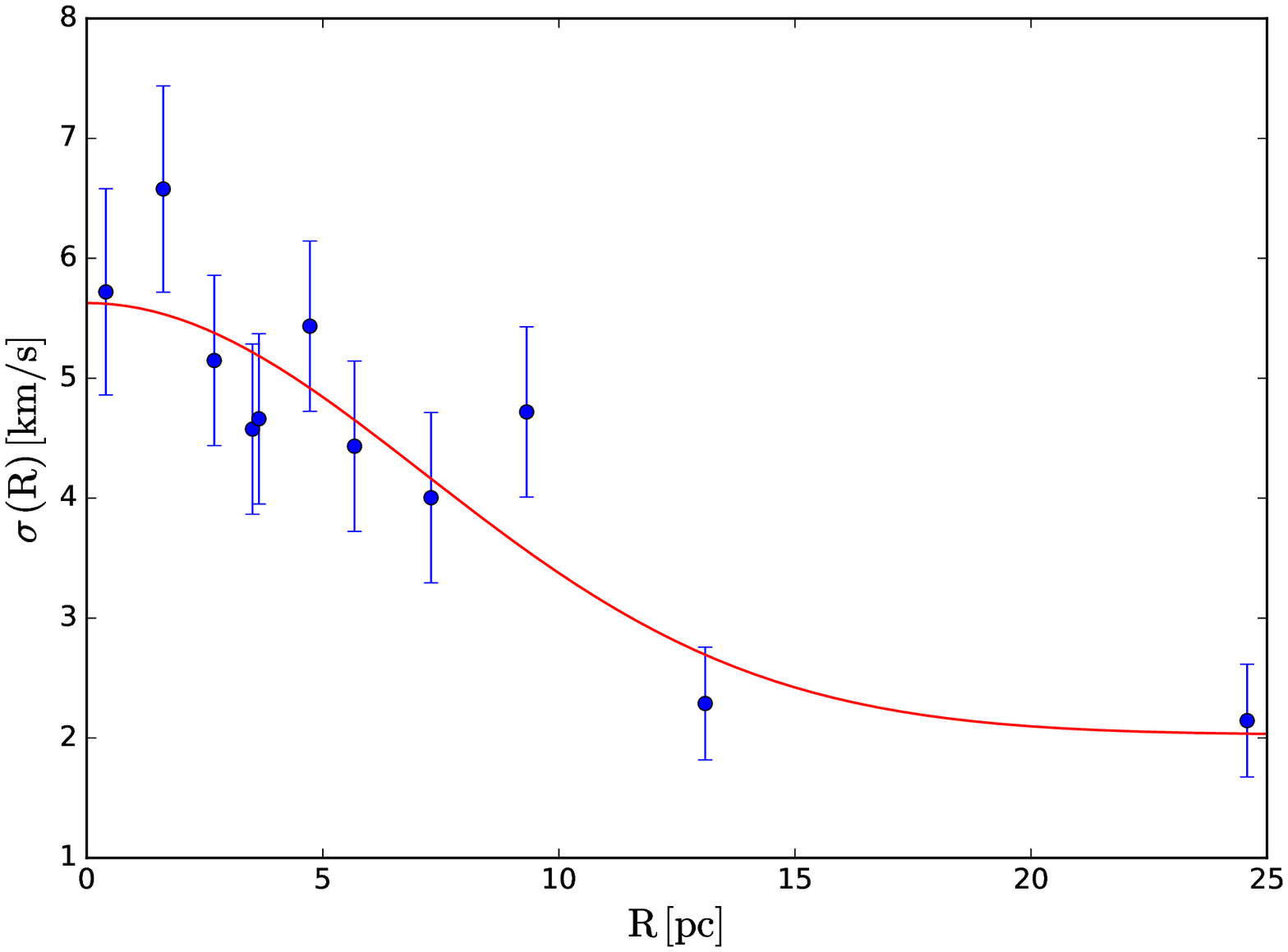}{0.3\textwidth}{NGC 6752}
          \fig{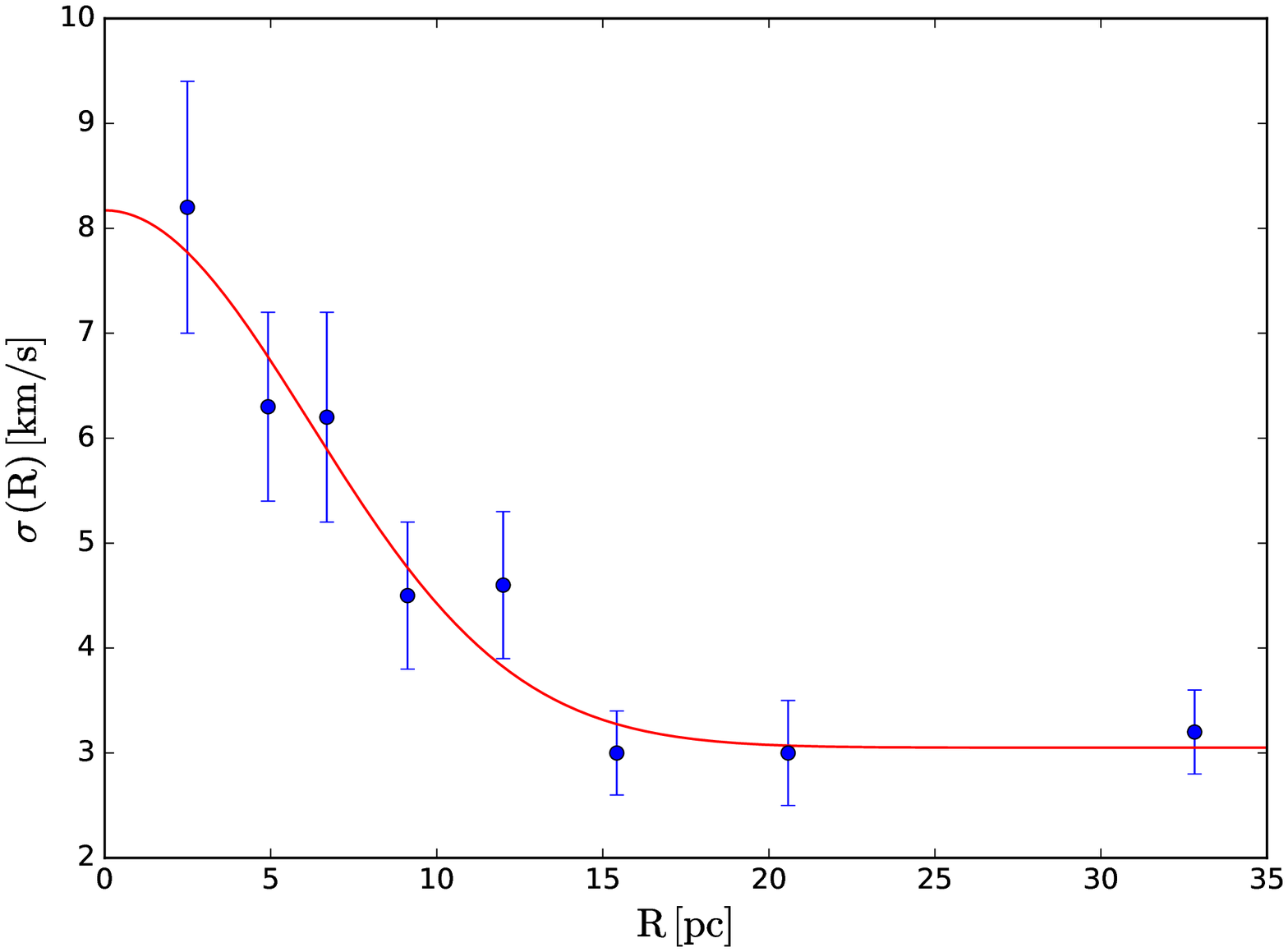}{0.3\textwidth}{NGC 7087}
          }          
          \caption{Projected velocity dispersion profiles for the 12 Galactic globular clusters of our sample, as a function
of radial distance in the system, with vertical error bars
showing the 1$\sigma$ dispersion at each radial bin. The solid curve
gives the best fit to the universal profile proposed.}
\end{figure*}

%
%
%
%
%
%

\begin{figure*}
\gridline{\fig{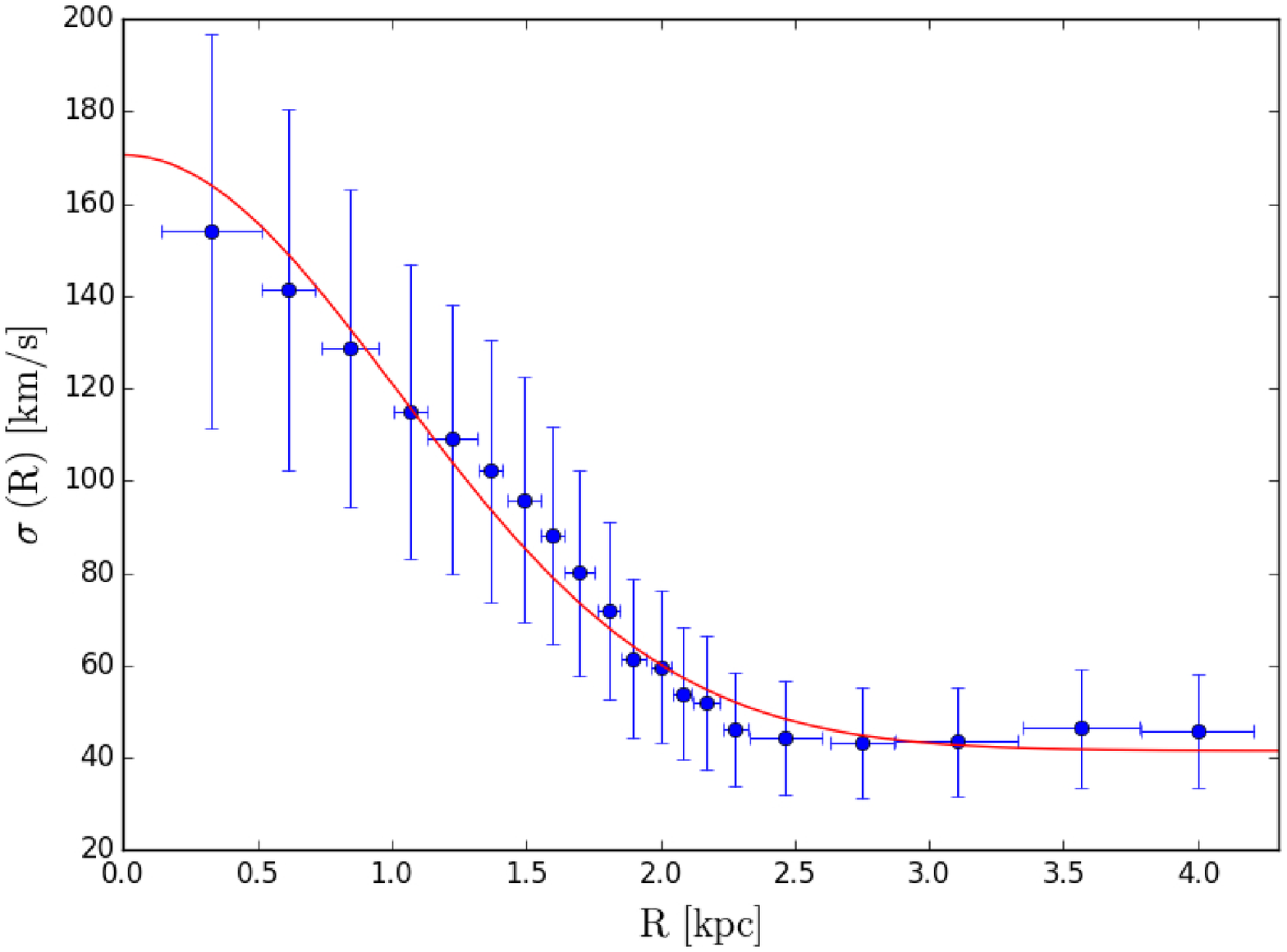}{0.3\textwidth}{NCG 0731}
          \fig{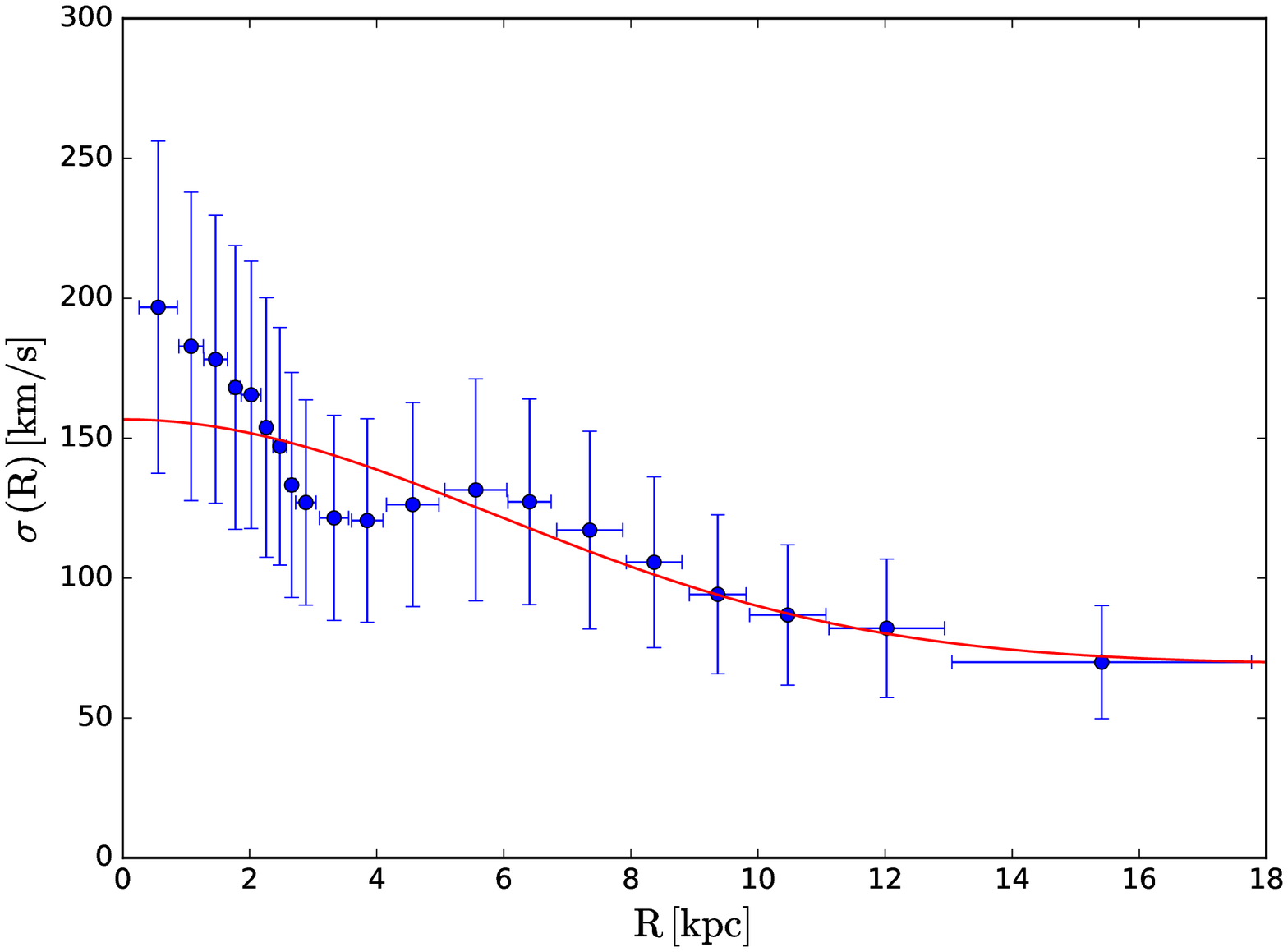}{0.3\textwidth}{NGC 0155}
          \fig{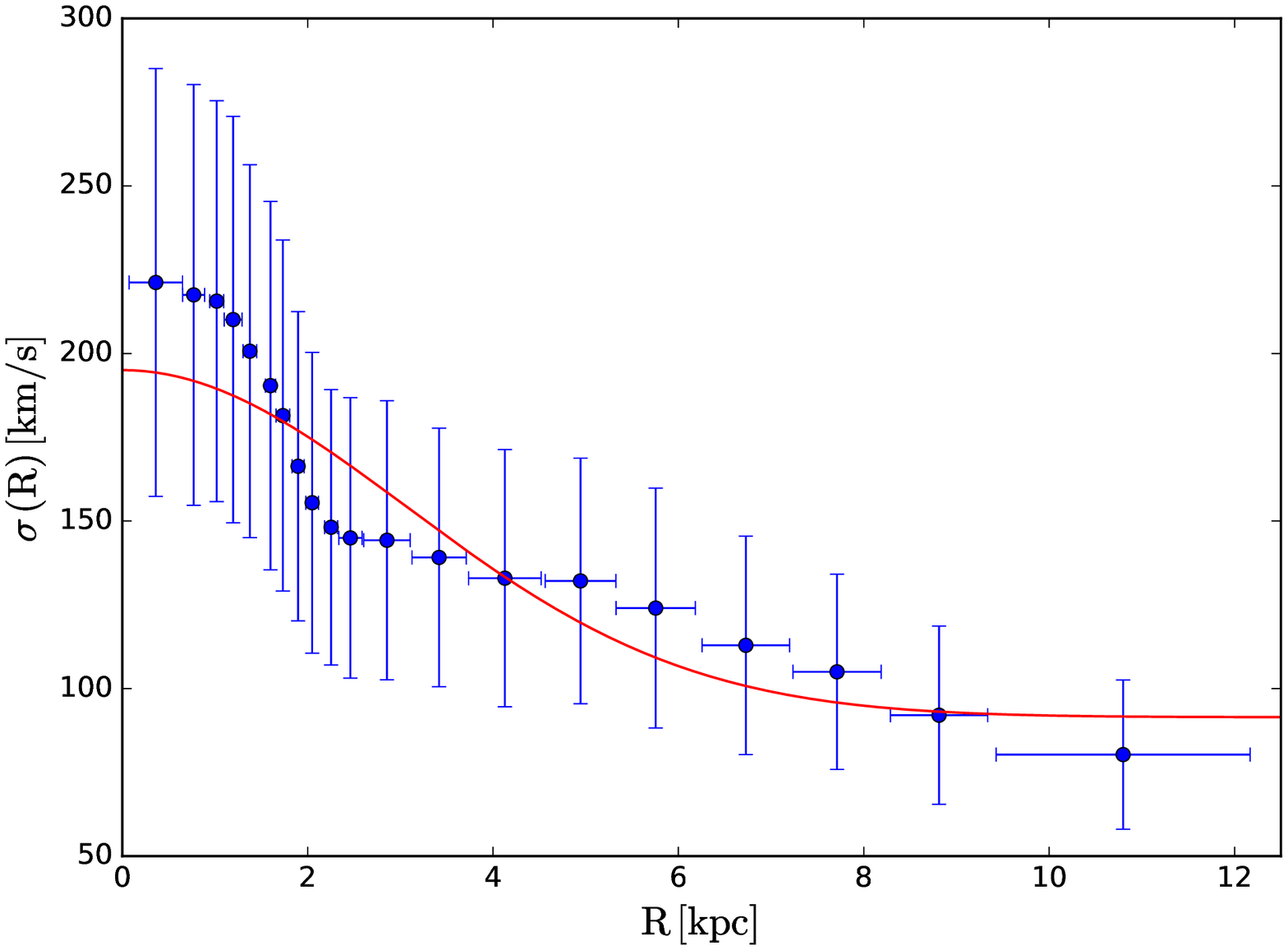}{0.3\textwidth}{NGC 0677}
          }
\gridline{\fig{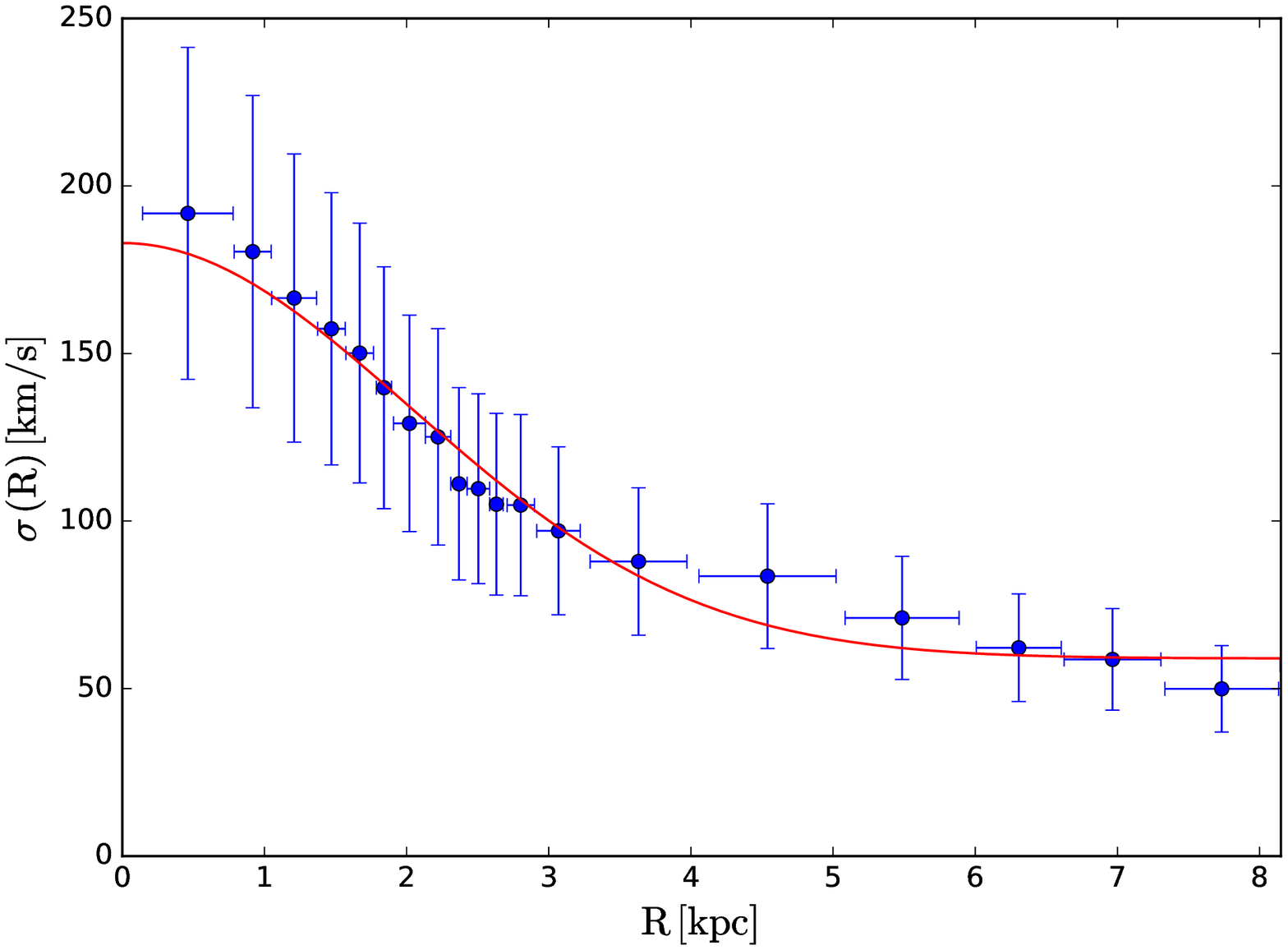}{0.3\textwidth}{NCG 0938}
          \fig{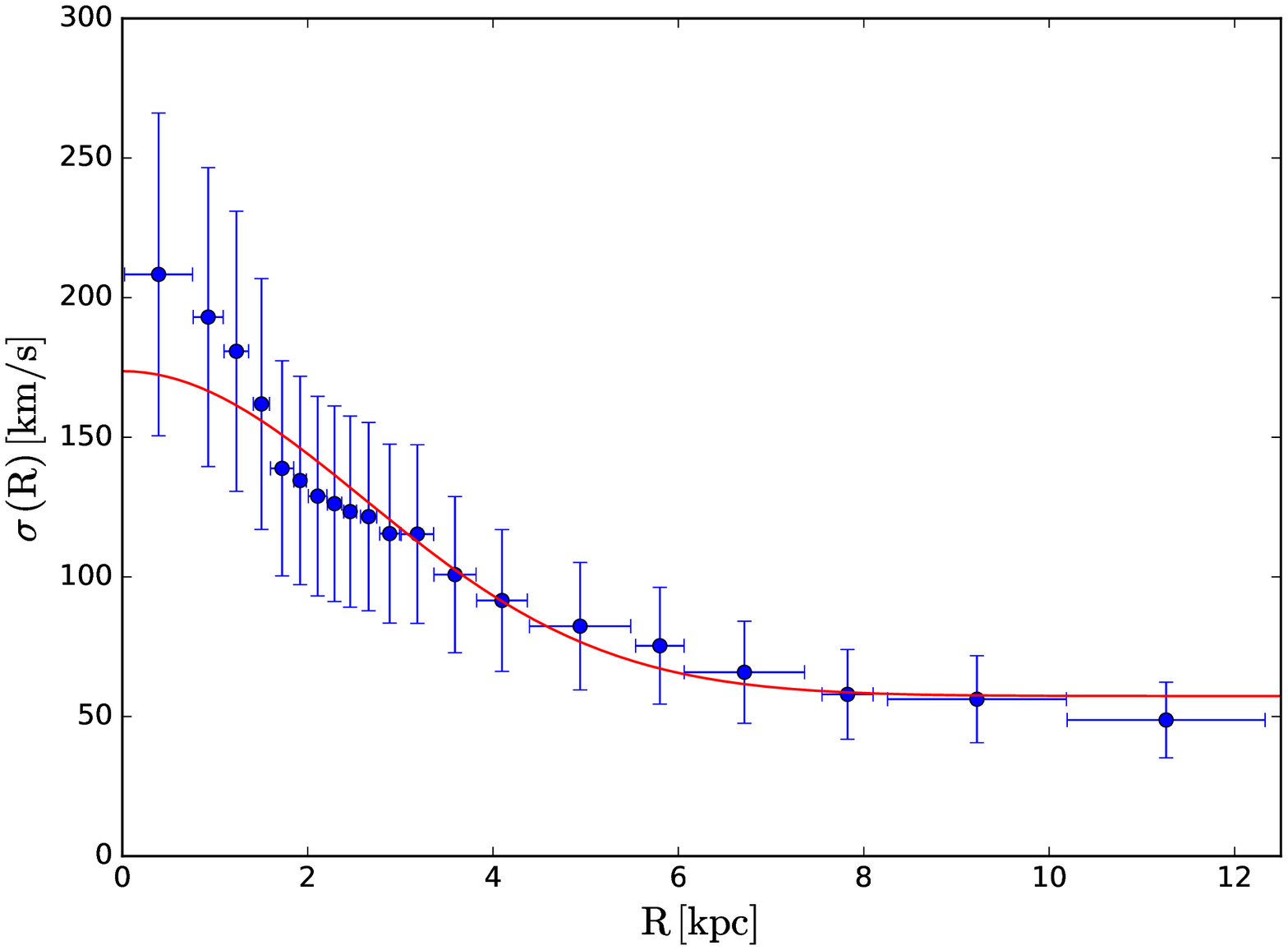}{0.3\textwidth}{NGC 0962}
          \fig{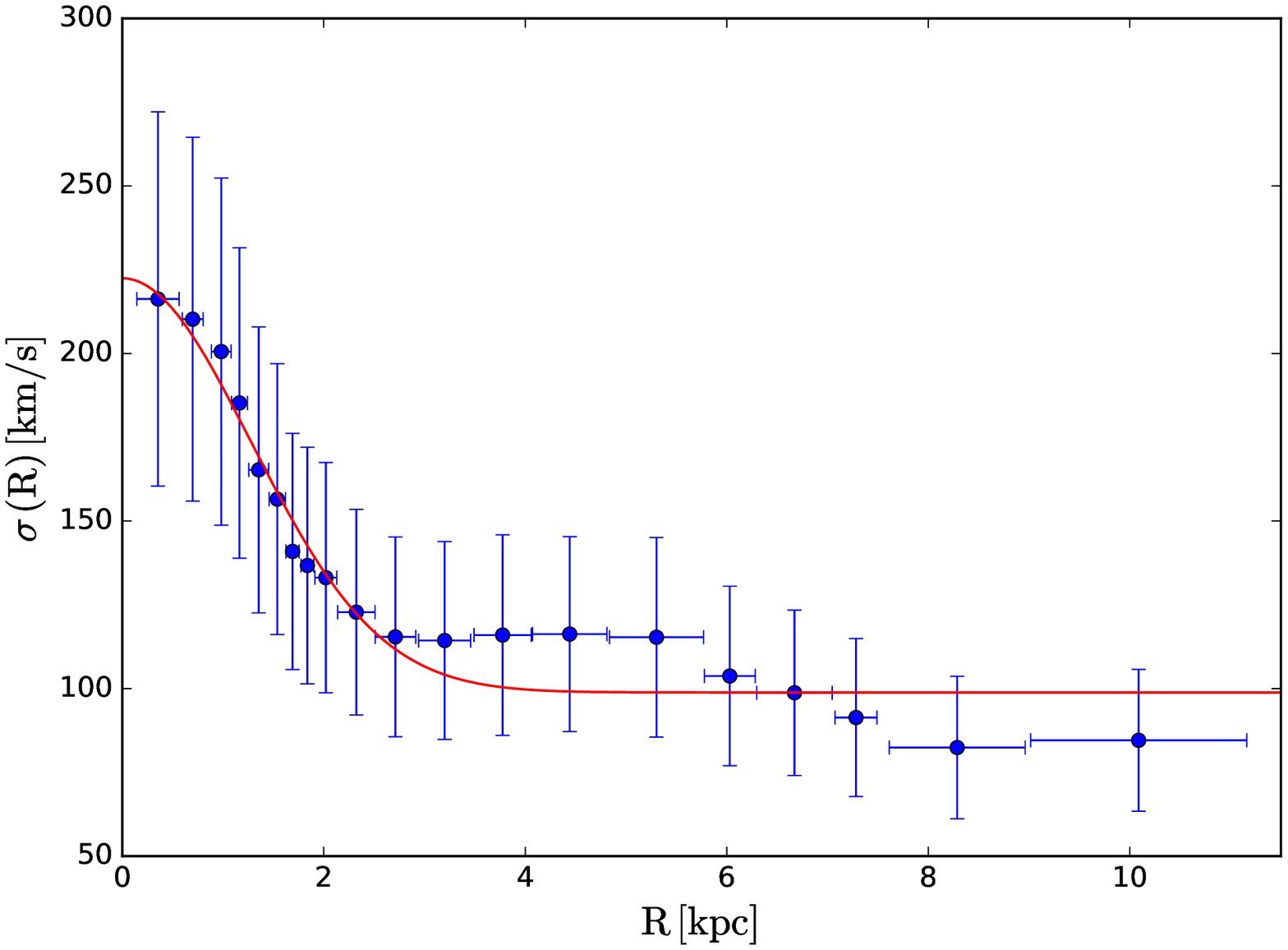}{0.3\textwidth}{NGC 1026}
          }
\gridline{\fig{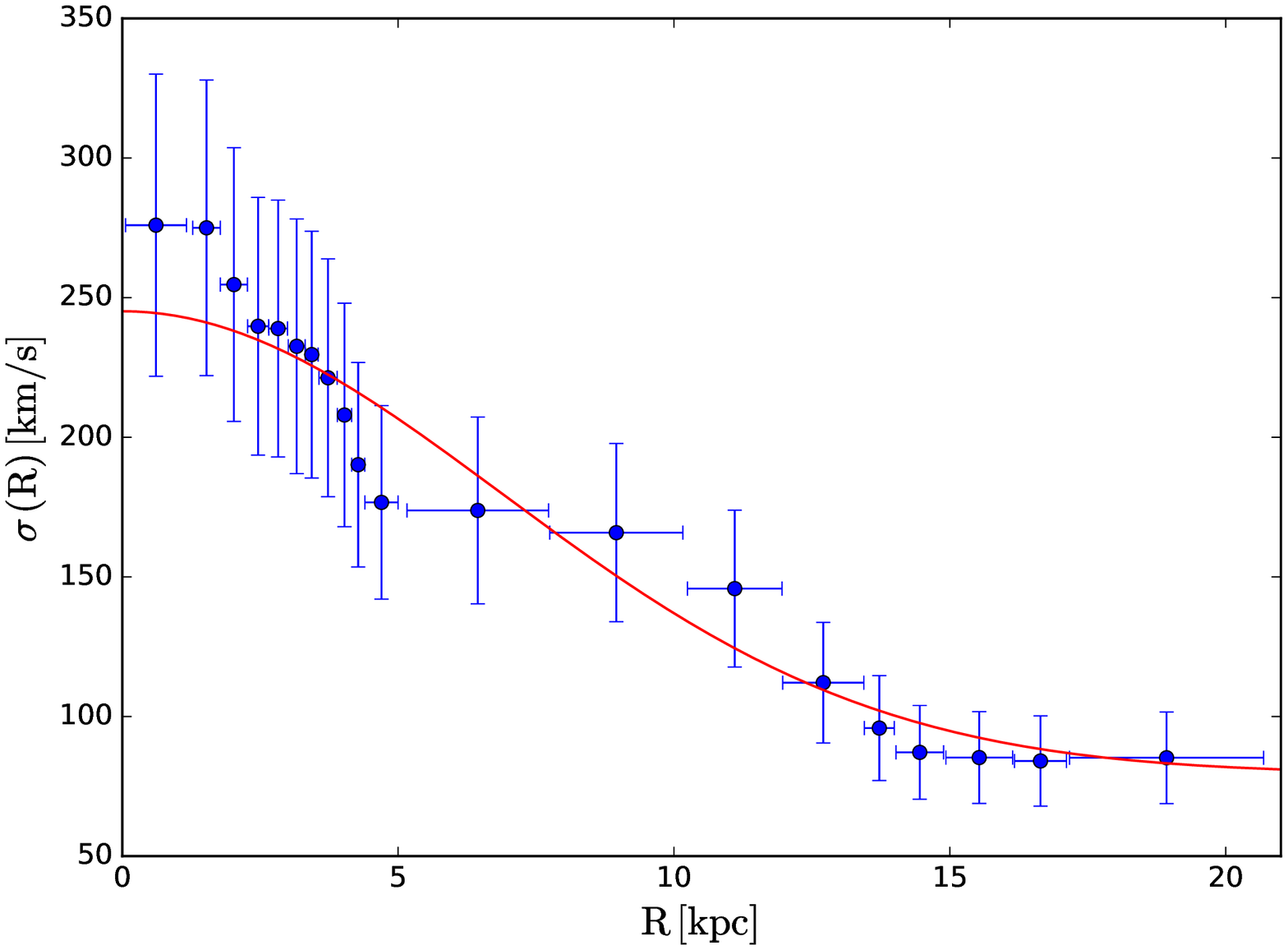}{0.3\textwidth}{NCG 1060}
          \fig{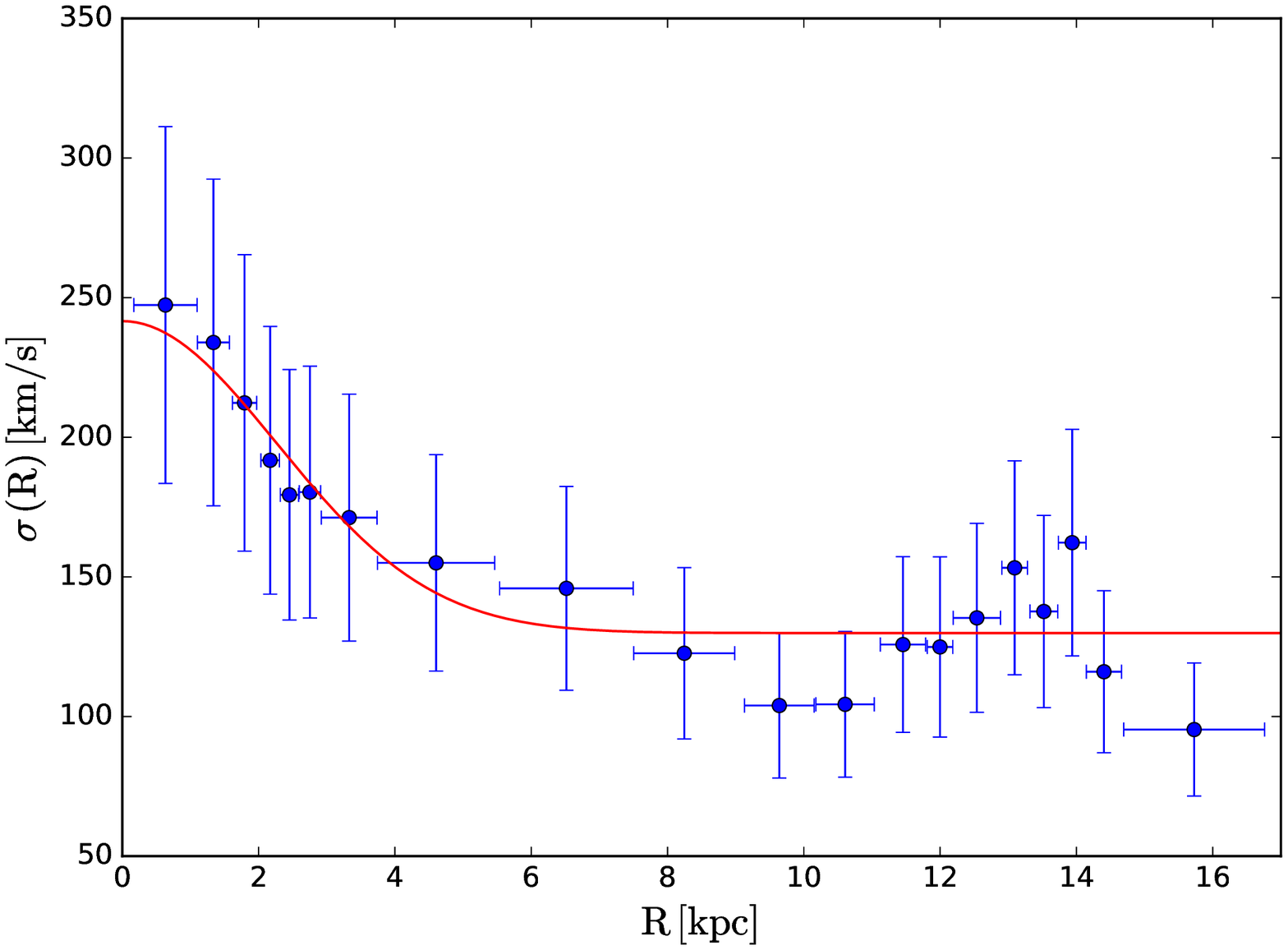}{0.3\textwidth}{NGC 4841A}
          \fig{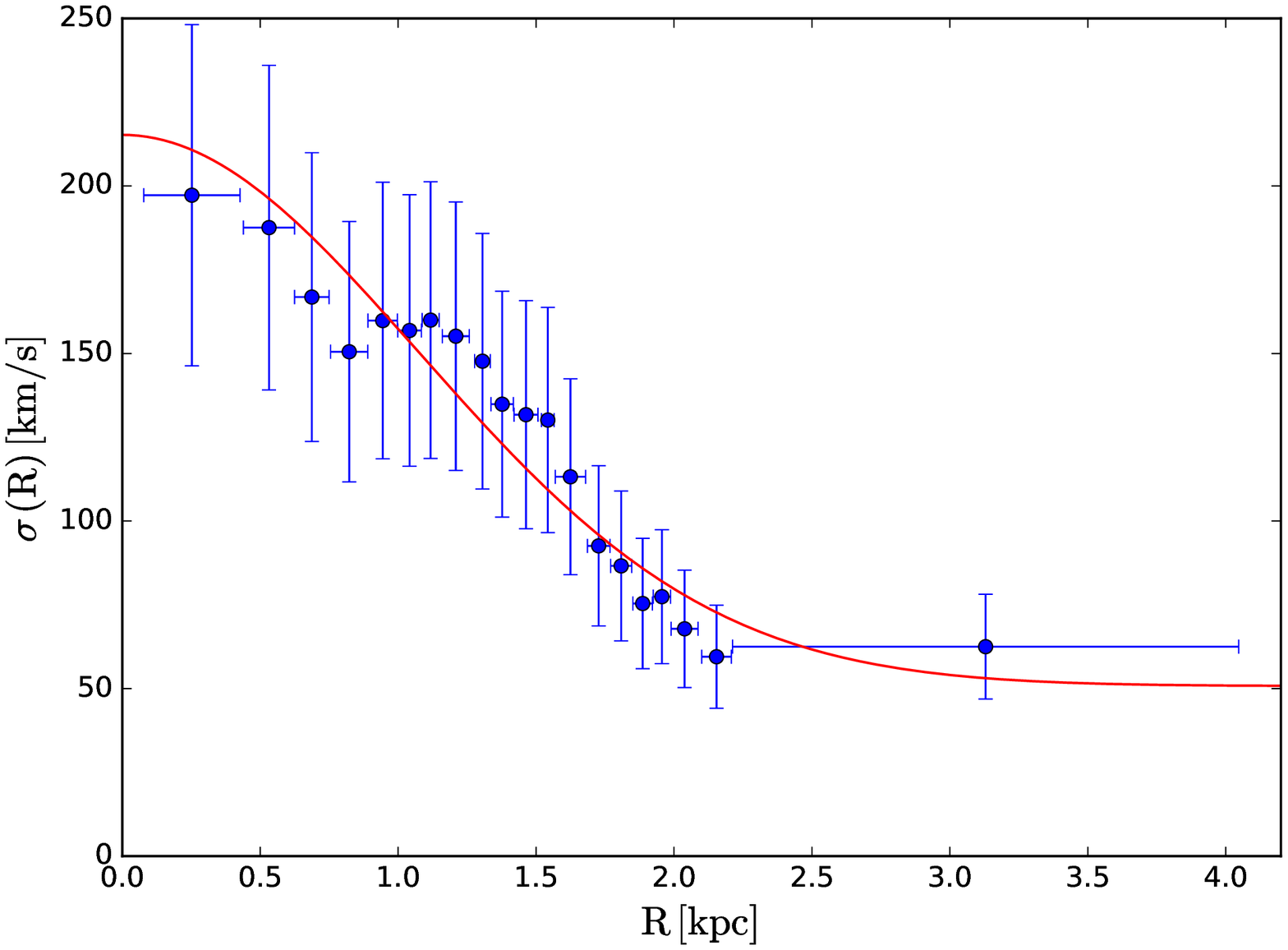}{0.3\textwidth}{NGC 5198}
          }
\gridline{\fig{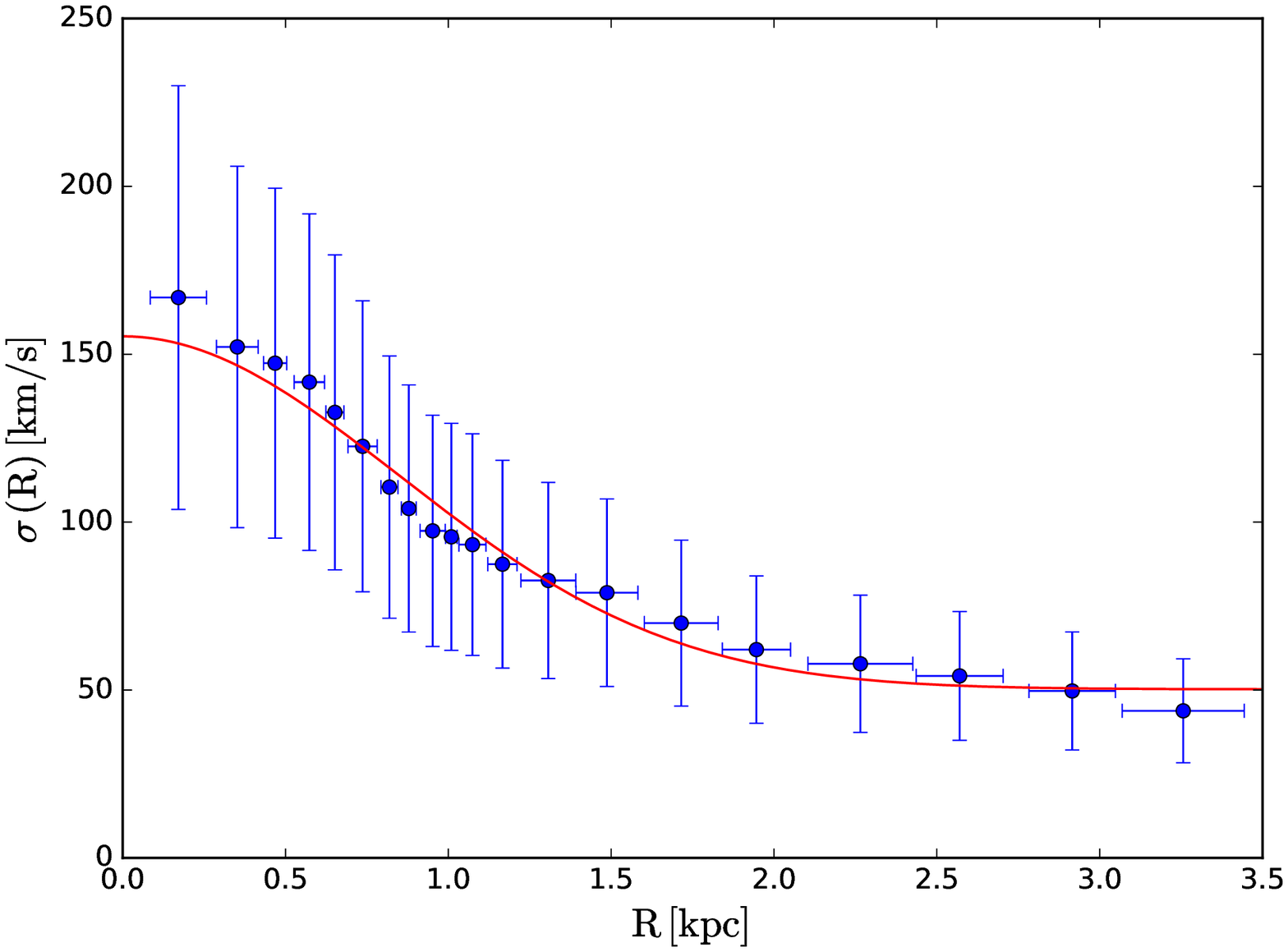}{0.25\textwidth}{NCG 5216}
          \fig{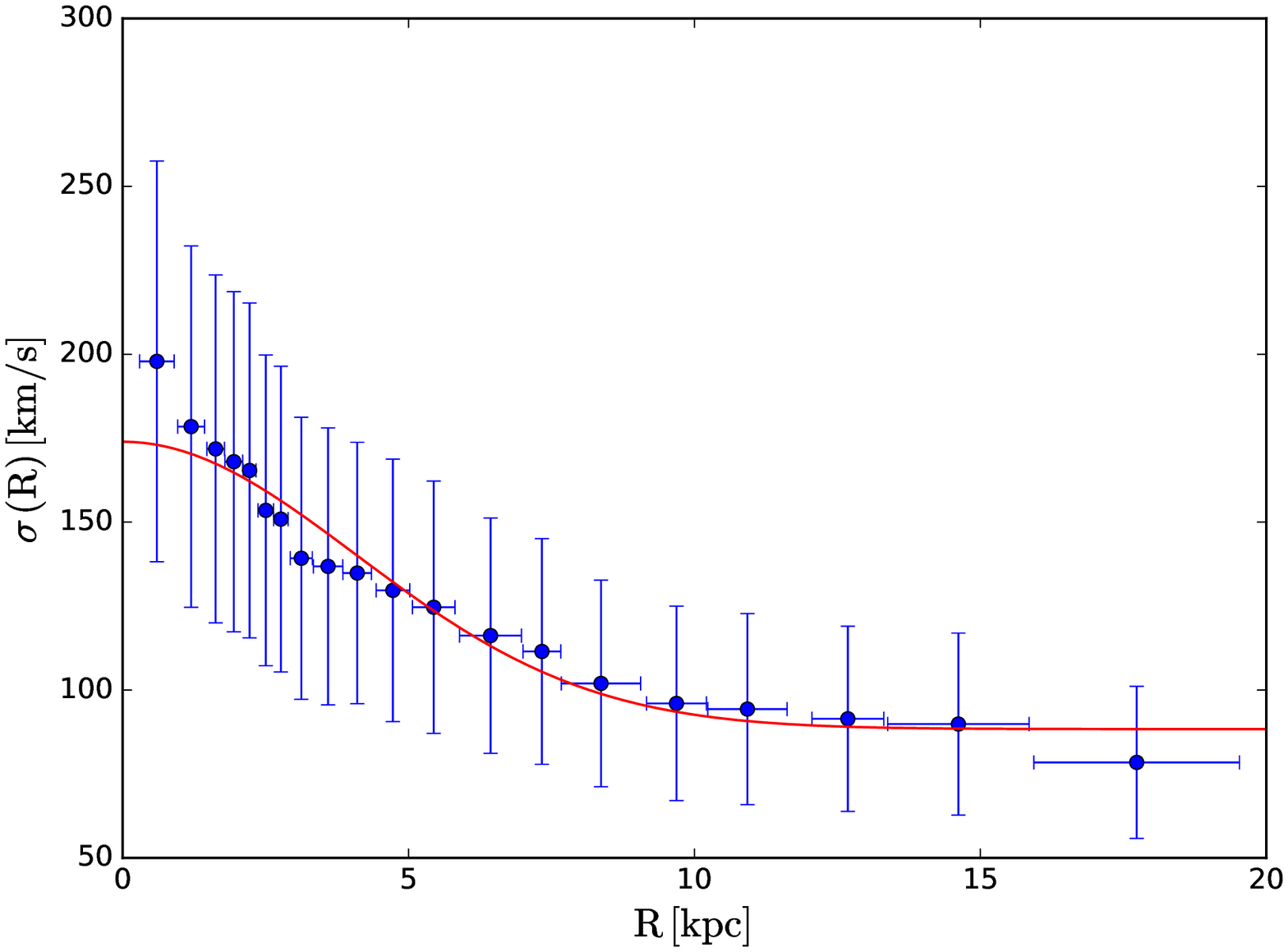}{0.25\textwidth}{NGC 6515}
          \fig{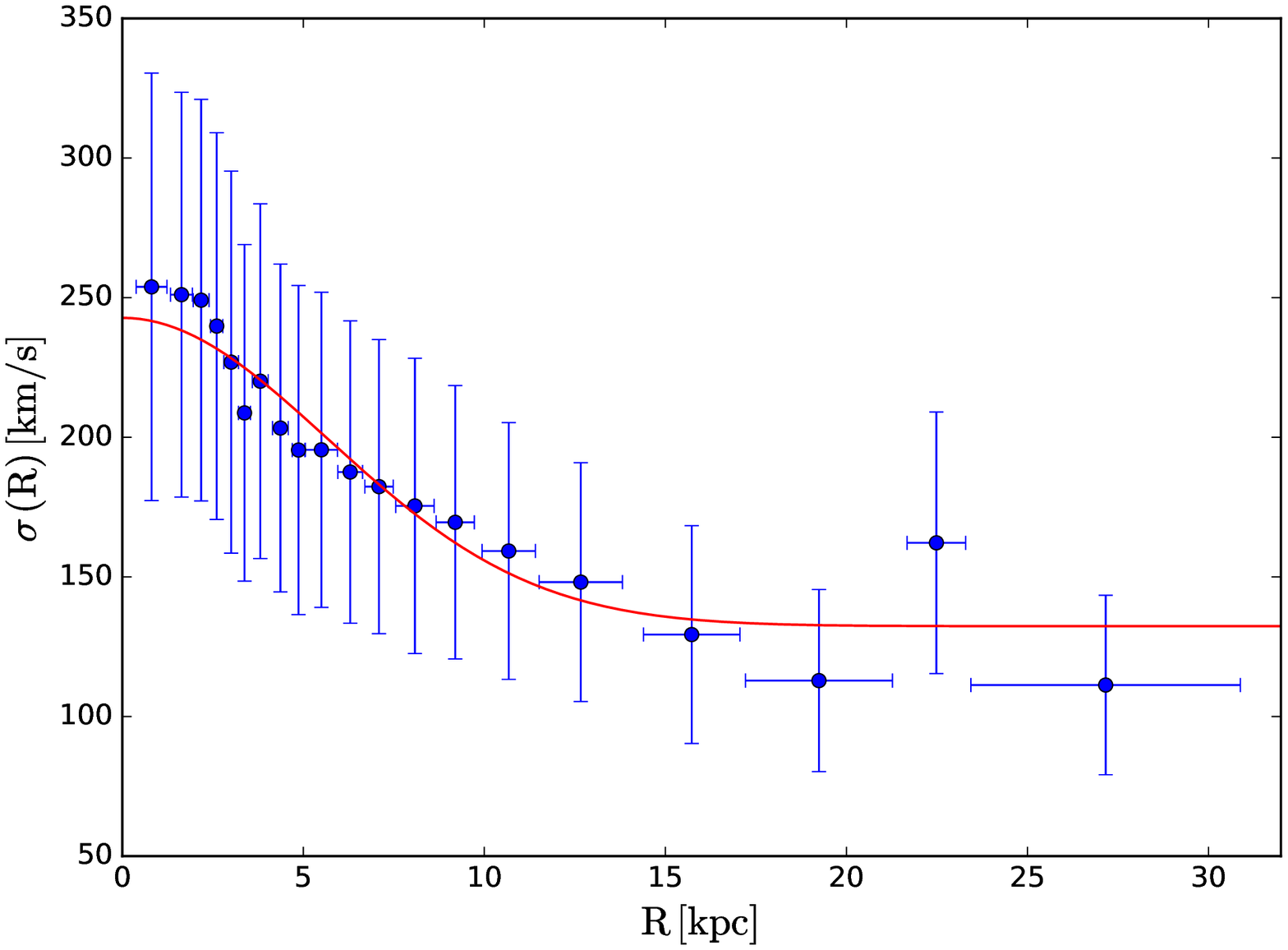}{0.25\textwidth}{NGC 7194}
          \fig{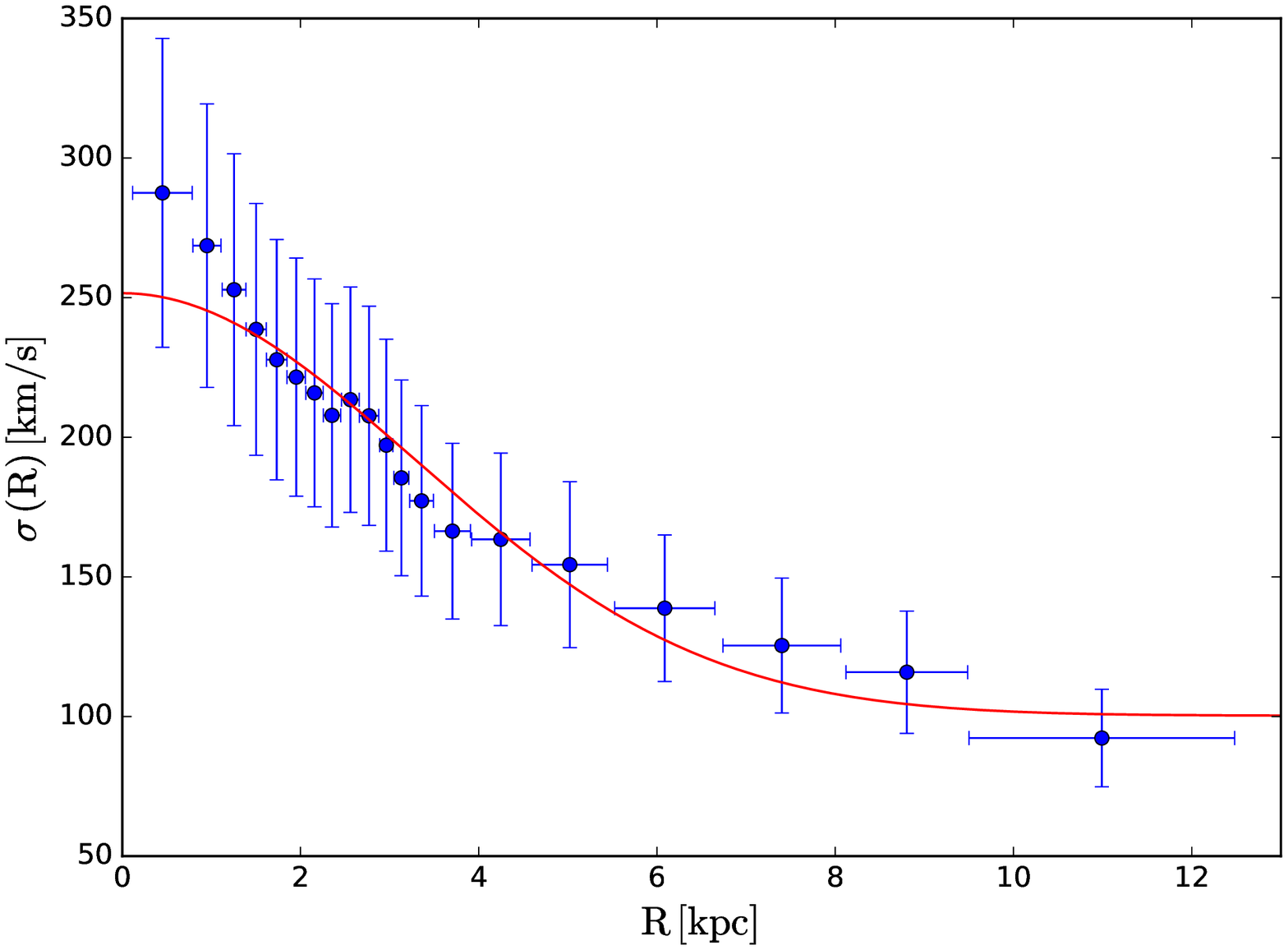}{0.25\textwidth}{NGC 7619}
          }          
          \caption{Projected velocity dispersion profiles for the 13 elliptical of our sample, as a function
of radial distance in the system, with vertical error bars
showing the 1$\sigma$ dispersion at each radial bin. The solid curve
gives the best fit to the universal profile proposed.}
\end{figure*}

\section{Empirical velocity dispersion profiles}

\subsection{Galactic globular clusters}

In data recently available from the Scarpa et al. group and the Lane et al. group, not only the central values, but 
the radial projected velocity dispersion profiles for several systems have been 
published, extending out to several half-light radii. As first mentioned by the Scarpa et al. group,
a flattening of the velocity dispersion profiles for Galactic globular clusters appears on crossing the $a_{0}$
acceleration threshold, in accordance with
MONDian gravity expectations. In Hernandez \& Jimenez (2012) some of us recently modeled 
the above globular cluster data to construct self-gravitating models for the systems studied, where both the projected 
surface brightness profiles and the observed velocity dispersion profiles have been accurately reproduced, using a modified 
gravity force law of the type described in section (2). Further, in Hernandez et al. (2013a) we showed that the asymptotic
velocity dispersion values for the clusters treated actually match the expected scaling of 
$\sigma_{\infty} \approx (GM a_{0})^{1/4}$ of MONDian gravity, through comparing with stellar synthesis models tuned to 
the ages and metallicities of each of the individual globular clusters studied. In the above works we also showed that
an empirical projected velocity dispersion profile of the form:

\begin{equation}
\sigma(R)= \sigma_{0} e^{-(R/R_{\sigma})^{2}} + \sigma_{\infty}
\end{equation}

\noindent serves to accurately model the reported velocity dispersion profiles of Galactic globular clusters.
In Equation (6) $\sigma_{\infty}$ is the asymptotic value for the system's velocity dispersion, the central value for this
quantity is given by $\sigma(0)=\sigma_{0}+\sigma_{\infty}$, and $R_{\sigma}$ gives a transition radius beyond which 
the asymptote is rapidly approached.

Here we include again the fits to the data of the Scarpa et al. and Lane et al. groups to Equation (6). We use a
non-linear least squares Levenberg-Marquardt algorithm to estimate the parameters, and obtain self-consistent $1\sigma$ 
confidence intervals, which are reported in Table 1. We have trimmed the total sample used in Hernandez et al. (2013a) 
to 12 globular clusters where the relative errors in the fitted parameters are in all cases smaller than 50$\%$. This excludes 4 globular 
clusters with poorly determined velocity dispersion profiles, also keeping the globular cluster sample from growing much 
beyond the numbers of available well measured systems in the category of low rotation elliptical 
galaxies. This last to ensure that the systematics relevant to any particular class of objects do 
not dominate and bias the overall comparisons we attempt. 

Figure 1 provides the observed velocity dispersion profiles for the Galactic globular
clusters of our sample with the best fit to Equation (6) for each system, showing clearly an excellent empirical
representation of the measured profiles. A clear decline
in the Newtonian inner region is evident, followed by a transition to a ``Tully-Fisher'' flat asymptote.

\subsection{Elliptical galaxies}

A sample of low rotation elliptical galaxies was obtained from S{\'a}nchez et al. (2016b),
in which 200 radial 2D velocity dispersion profiles were produced from the second data release
of the CALIFA survey (S{\'a}nchez et al. 2012; Walcher et al. 2014; S{\'a}nchez et al. 2016a)
using the PIPE 3D tool.
To ensure that the selected sub-sample was composed of systems with the least dynamical support besides velocity
dispersion, we selected galaxies based on their morphology (E0-E3). Our sample comprises only extremely slow rotators,
with an average value of a maximum rotation velocity to certral sigma per galaxy of $V_{max}/\sigma_{0} =0.213$, very early
type systems with negligible total gas content.

We use data directly obtained from S\'anchez et al. (2016b), consisting of 2D radial stellar 
velocity dispersion observations with their respective uncertainties, as well as associated flux for a correct weighted 
statistical development. From the 2D velocity dispersion observations we extract a projected radial velocity dispersion 
representation. 
Due to the large number of pointings in each galaxy, we averaged the projected velocity dispersion
observations within 20 radial bins (Figure 2), each with a corresponding 1$\sigma$ dispersion which
greatly overshadows the intrinsic observational error for each data point. 
In spite of the high 1$\sigma$
dispersion towards the center of the galaxies,
we adopt this method for statistical consistency with the globular clusters treated.
Monte Carlo generated errors for each data point are only about 15 km/s,
while the radial bin average confidence intervals can reach up to 50 km/s.
As described in Sanchez et al. (2016a,b)
 velocity dispersion measurements of under 40 ${km/s}$ from the 
middle resolution setup V1200  are
not reliable, so we have discarded these data points in our estimates.
Using again a non-linear least squares Levenberg-Marquardt algorithm, we fit the proposed universal 
function of Equation (6) to the projected velocity dispersion profiles and obtain optimal parameters as well as their 
respective confidence intervals. After discarding various systems with poorly determined parameters, with fractional errors
larger than $50\%$ in the fitted parameters, we compile a final sample of 13 elliptical galaxies, as reported in Table 1.

Figure 2 shows the radial binning of the projected 2D radial velocity dispersion of the elliptical galaxies in the sample, 
as well as the best fit to the universal function of Equation (6). Note again the suitability of the
fits in all cases.

\begin{figure}
\hskip -10pt \includegraphics[width=8.8cm,height=6.8cm]{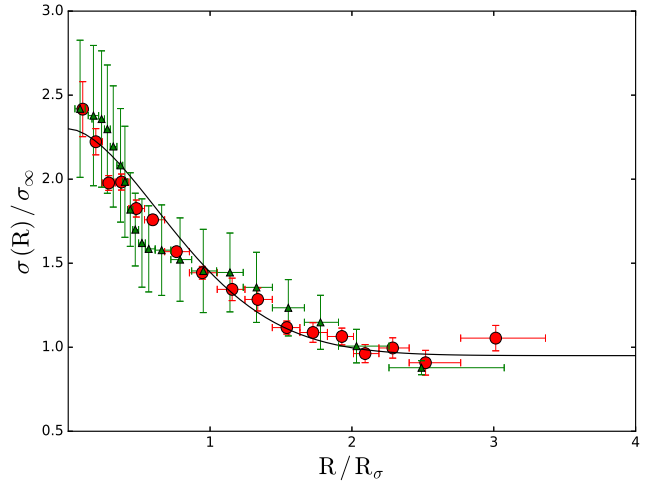}
\caption{Observed projected velocity dispersion profiles in units of the corresponding
$\sigma_{\infty}$ values as a function of radius normalized to the corresponding $R_{\sigma}$ values, for
two distinct systems. Red circles correspond to measurements for the Galactic globular cluster NGC 5139, with a
central velocity dispersion of $15.2 km/s$ and $R_{\sigma}=16.7 pc$. Green triangles correspond to the elliptical
galaxy NGC 0677, with a central velocity dispersion of $195 km/s$ and $R_{\sigma}= 4.3 kpc$. The consistency of the
two scaled projected velocity dispersion profiles is evident, despite the range of scales covered by the comparison.}
\end{figure}

\begin{figure}
\hskip -10pt \includegraphics[width=8.8cm,height=6.8cm]{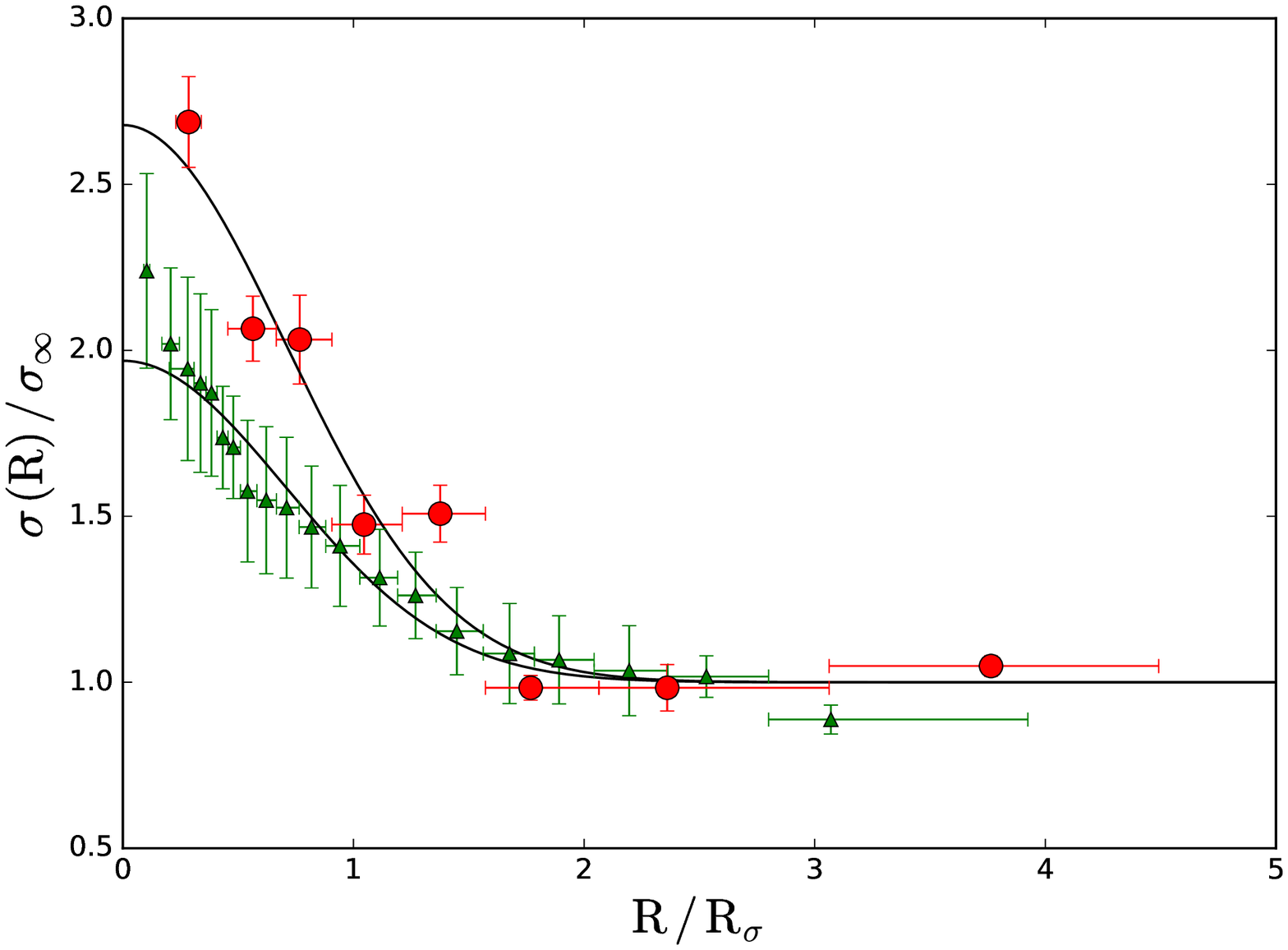}
\caption{Observed projected velocity dispersion profiles in units of the corresponding
$\sigma_{\infty}$ values as a function of radius normalized to the corresponding $R_{\sigma}$ values, for
two distinct systems. Red circles correspond to measurements for the Galactic globular cluster NGC 7078, with a 
central velocity dispersion of $8.1 km/s$ and $R_{\sigma}=8.7 pc$. Green triangles show the elliptical
galaxy NGC 6515, with a central velocity dispersion of $174 km/s$ and $R_{\sigma}= 5.8 kpc$. The two scaled profiles can
be modeled through the same functional form of the proposed
universal profile, the Figure also shows the tendency for the scaled profiles of elliptical galaxies to be flatter and 
for globular clusters to be steeper.}
\end{figure}




\begin{table*}
\begin{flushleft}
  \caption{Parameters for velocity dispersion profiles of the the pressure supported systems.}
  \begin{tabular}{@{}llllll@{}}
  \hline
 \hline
   Type of system & Identification \,\, & $\sigma_{0} (km/s) $ & $\sigma_{\infty} (km/s)$  & $R_{\sigma} (pc)$ \\
 \hline
 Globular Cluster & NGC 0104       & $4.5 \pm 0.4$   & $5.0 \pm 0.1$ & $14.8 \pm 0.9$ \\
 Globular Cluster & NGC 1851      & $4.1 \pm 1.7$   & $3.9 \pm 0.4$ & $7.9  \pm 2.3$ \\
 Globular Cluster & NGC 1904      & $2.4 \pm 0.6$   & $2.0 \pm 0.4$ & $10.3 \pm 3.3$ \\
 Globular Cluster & NGC 2419      & $4.7 \pm 0.8$   & $1.0 \pm 0.4$ & $1.8  \pm 0.3$ \\
 Globular Cluster & NGC 5139      & $8.0 \pm 0.7$   & $7.2 \pm 0.4$ & $16.7 \pm 1.9$ \\
 Globular Cluster & NGC 6171      & $1.4 \pm 0.4$   & $2.7 \pm 0.2$ & $3.8  \pm 1.1$ \\
 Globular Cluster & NGC 6218      & $3.3 \pm 0.9$   & $1.4 \pm 0.5$ & $4.7  \pm 1.2$ \\
 Globular Cluster & NGC 6341      & $3.8 \pm 0.7$   & $3.1 \pm 0.4$ & $6.9  \pm 1.4$ \\
 Globular Cluster & NGC 6656      & $3.8 \pm 1.3$   & $3.3 \pm 0.8$ & $7.4  \pm 1.7$ \\
 Globular Cluster & NGC 6752      & $3.5 \pm 0.8$   & $2.0 \pm 0.3$ & $10.7 \pm 1.4$ \\
 Globular Cluster & NGC 7078      & $5.1 \pm 1.0$   & $3.0 \pm 0.3$ & $8.7  \pm 1.7$ \\
 Globular Cluster & NGC 7099      & $1.8 \pm 0.5$   & $2.2 \pm 0.3$ & $6.2  \pm 2.3$ \\
 \hline 
 Elliptical Galaxy & NGC 0155     & $87.6 \pm 25.5$   & $69.1 \pm 25.4$  & $(8.4\pm 3.6)\times 10^{3}$   \\
 Elliptical Galaxy & NGC 0677     & $103.6 \pm 24.4$    & $91.4 \pm 16.7$  & $(4.3\pm 1.6)\times 10^{3}$  \\
 Elliptical Galaxy & NGC 0731     & $129.0 \pm 26.9$   & $41.6 \pm 6.6$ & $(1.4\pm 0.2)\times 10^{3}$ \\
 Elliptical Galaxy & NGC 0938     & $124.0 \pm 25.8$   & $59.0 \pm 7.0$  & $(2.9\pm 0.5)\times 10^{3}$   \\
 Elliptical Galaxy & NGC 0962     & $116.3 \pm 22.5$    & $57.3 \pm 8.5$  & $(3.7\pm 0.8)\times 10^{3}$   \\
 Elliptical Galaxy & NGC 1026     & $123.6 \pm 37.7$   & $98.8 \pm 8.8$ & $(1.8\pm 0.4)\times 10^{3}$   \\
 Elliptical Galaxy & NGC 1060     & $165.6 \pm 18.8$   & $79.5 \pm 16.7$ & $(9.7\pm 2.0)\times 10^{3}$   \\
 Elliptical Galaxy & NGC 4841A     & $111.7 \pm 46.7$   & $129.9 \pm 8.9$  & $(3.2\pm 1.4)\times 10^{3}$   \\
 Elliptical Galaxy & NGC 5198     & $163.6 \pm 27.3$    & $53.8 \pm 14.7$  & $(1.5\pm 0.2)\times 10^{3}$   \\
 Elliptical Galaxy & NGC 5216     & $105.1 \pm 28.3$   & $50.2 \pm 9.6$  & $(1.2\pm 0.3)\times 10^{3}$   \\
 Elliptical Galaxy & NGC 6515     & $85.6 \pm 24.3$   & $88.4 \pm 13.5$  & $(5.8\pm 2.2)\times 10^{3}$   \\
 Elliptical Galaxy & NGC 7194     & $110.4 \pm 33.6$   & $132.3 \pm 18.7$  & $(8.0\pm 3.2)\times 10^{3}$   \\
 Elliptical Galaxy & NGC 7619     & $151.3 \pm 21.8$   & $100.3 \pm 14.7$  & $(4.6\pm 0.9)\times 10^{3}$   \\
 \hline

\end{tabular} 

The first two columns give the object class and identifier for the systems studied. The following three entries
show the parameters of the fits to the observed projected velocity dispersion profiles and their
confidence intervals. Data from the Scarpa et al. group and the Lane et al. group for Galactic globular clusters,
and data from the CALIFA collaboration for elliptical galaxies. 
\end{flushleft}
\end{table*}

Although the general fit of the proposed universal velocity dispersion profile is clearly very good across the range of objects treated (see the full figure set), specially towards the flattened outer regions, it is also clear that in a number of both globular clusters and elliptical galaxies, a slight central enhancement in the data with respect to the fitted function appears. This cuspy
velocity dispersion profiles, e.g. the globular cluster NGC 2419 and elliptical galaxies NGC 0962 and NGC 1060, could signal the presence of a central black hole.  This is a common feature in early type galaxies, and could also appear in globular clusters. For example, Baumgardt (2017) recently concluded from dynamical studies of NGC 5139 that this cluster likely harbors a 40,000 solar mass black hole in its center. From Figure 1 we see that indeed, this particular system shows the most prominent central velocity dispersion excess with respect to the universal profile fitted of all the globular clusters treated.

Fortunately, this inner velocity dispersion excesses are very central features which do not affect the determinations of $R_{\sigma}$ or $\sigma_{\infty}$, the parameters which can be associated to MONDian dynamical predictions.

\section{Comparisons with MONDian expectations}

Comparing Figures 1 and 2, we see that although the physical scale is changing from pc to kpc,
the velocity profiles of globular clusters and ellipticals can both be fitted adequately by the same family
of functions. Further, while the peak $\sigma_{0}$ is on average different for globular clusters and ellipticals when
scaled with $\sigma_{\infty}$, the transition from a declining velocity profile to a flat one emerges with
remarkable consistency at a scaled radius of $R/R_\sigma \sim 2$.

From Equation (6) we see that the central velocity dispersion values, given by $\sigma(0)=\sigma_{0}+\sigma_{\infty}$, span two
orders of magnitude, with the scale radii $R_{\sigma}$ covering three orders of magnitude, and the total baryonic masses
going from a few $10^{4} M_{\odot}$ scale for the smallest Galactic globular clusters to  $10^{11} M_{\odot}$
for the elliptical galaxies. If we now scale the projected 
velocity dispersion profiles of the various systems presented in the previous section, and consider
$\sigma(R)/\sigma_{\infty}$ as a function of $R/R_{\sigma}$ for each profile, only one degree of freedom remains,
given by the central velocity dispersion, $\sigma(0)/\sigma_{\infty}$.

Therefore, if we choose distinct systems having equal $\sigma(0)/\sigma_{\infty}$ values, they should look identical in a
$\sigma(R)/\sigma_{\infty}$ vs. $R/R_{\sigma}$ plot. Figure 3 shows the scaled velocity dispersion profiles of two different
systems chosen to have very similar $\sigma(0)/\sigma_{\infty}$ values. Circles correspond to the Galactic globular cluster
NGC 5139, while triangles represent data for the elliptical galaxy NGC 0677.
The solid line shows a fit to Equation (6) for the combined data set, the
self-similarity of the two quite distinct pressure supported systems is
evident; once scaled, both observed velocity dispersion data samples are
consistent with each other to within their respective confidence intervals.
Actual physical
radial scales vary from $R_{\sigma}=16.7 pc$ of the globular cluster shown to $R_{\sigma}=4.3 kpc$ of the elliptical
galaxy.  We see pressure supported systems as showing a simple 
empirical kinematic profile across many orders of magnitude in scale and distinct classes of systems.

Despite the overlap evident in Figure 3, typically we find velocity dispersion profiles for elliptical galaxies to be rather
flat, while those of Galactic globular clusters tend to be steeper. This is illustrated in Figure 4, which is analogous 
to Figure 3, but shows more typical examples for the two astrophysical systems analyzed, the Galactic globular
cluster NGC 7078 and the elliptical galaxy NGC 6515. Although the same functional form is evident, 
the degree of central concentration clearly varies as mentioned above.


Regarding Galactic globular clusters, it has been suggested that the
flattening of the velocity dispersion profiles could be due to standard
Newtonian physics, in particular tidal heating from the overall Milky
Way potential (K\"upper 2010). Fortunately, 10 of the
globular clusters included in our study form part of a wider sample
for which not only distances but also well measured proper motions
are available. Indeed, Allen et al. (2006) and Allen et al. (2008)
calculated orbital properties for a large sample of 54 Galactic
globular clusters within a fully consistent Newtonian galactic model
including baryonic disk, bulge and stellar halo, as well as a dominant
dark matter halo. In those studies, the authors calculate detailed tidal
radii for all globular clusters, not under any 'effective mass'
approximation, but through fully calculating the derivative of the total
Galactic gravitational force, including evaluating the gradients in
acceleration across the extent of the clusters, along the entire
orbital trajectories.

In Hernandez et al. (2013a) one of us used fits to Equation (6) to
obtain $R_{\sigma}$ for Galactic globular clusters including the ones
in our present study, with the exception of NGC 2419, to compare with
the Newtonian tidal radii of Allen et al. (2006) and Allen et al.
(2008). For the globular clusters in our present study (excluding NGC
2419) the ratio of the Newtonian tidal radii - at peri-Galacticon - to
$R_{\sigma}$ ranges from between 3 to 14.7, with a mean value of 7.1.
Thus, it appears extremely unlikely than Newtonian tides are responsible
for the flattening observed. The case of NGC 2419 is also clear, proper
motions have only recently become available due to the extreme distance
at which it finds itself, Massari et al. (2016) provided the first
proper motion studies for this cluster, and orbital integration within
a Newtonian Galactic potential to show its galactocentric radius
oscillates between 53 and 98 kpc, making it highly insensitive to
Galactic tides. In going to MONDian schemes, as the gravitational force
falls with $R^{-1}$ rather than $R^{-2}$, the gradient of it is much
reduced, and Galactic globular clusters become more robust to Galactic
tides than under Newtonian dynamics (e.g. Hernandez \& Jimenez 2012),
justifying their treatment here as isolated systems.

For the case of the elliptical galaxies in the sample, we use two different 
criteria to determine interaction and isolation. The first comes directly from
the morphological classification of the CALIFA 3rd data release, which
is essentially a visual examination of the optical images of each object
made by five different CALIFA members, where signs of mergers or
interactions were sought. All our sample galaxies except one, NGC 4841A,
appear as isolated under this criterion.
In a second test, the $f$ parameter discussed
in Varela et al. (2004) is calculated for all CALIFA galaxies (Walcher et al. 2014),
which is an estimate of the logarithmic ratio between inner and tidal forces acting upon each galaxy,
given in terms of the projected distance between target and neighbor galaxies, the size
of the target galaxy
and the ratio of their masses as estimated from total luminosities assuming a constant
mass-to-light ratio.
Galaxies with $f$ values lower than $-4.5$ are considered completely isolated,
while galaxies with $f<-2$ are considered as not affected by tidal forces
of their neighboring galaxies. Again, all the galaxies from our sample except for
NGC 4841A show $f$ values lower than $-2$, indicating no tidal effects are evident upon them.
In the case of NGC 4841A, which shows clear visual signs of interaction and
has a value of $f=-0.01$, a distinct perturbation in its velocity dispersion profile
observations can be observed at around 12-14 kpc. The fit shown includes this slightly perturbed
region, indicating the robustness of the fit to details; excluding all data beyond
12 kpc yields fit parameters well within the reported confidence intervals of
those resulting from using the full data sample.

Other Newtonian explanations for the flattening in the globular cluster
velocity dispersion profiles have been put forward in the literature,
e.g. Kennedy (2014) showed that internal N-body relaxation processes
can yield results consistent with observations. However, under any
Newtonian explanation, it would then be down to a curious coincidence
that the amplitude of $\sigma_{\infty}$ should scale with the total
mass of the clusters in question (as inferred from detailed population
synthesis models tuned to the particular ages and metallicities of each
from McLaughlin \& van der Marel 2005) in consistency with MONDian
"Tully-Fisher" expectations, Equation (3), as shown in Hernandez et al.
(2013a).

We end this section with Figure 5, which contains the principal result of our study. The Figure shows $R_{\sigma}$
and $\sigma_{\infty}$ values for the fits to both of the pressure supported systems treated, in a logarithmic
plane. The systems shown cover a range of $1.4 - 165.6 km/s$ in central velocity dispersion and $1.8pc - 9.7kpc$ 
in the scale radius $R_{\sigma}$. The two distinct classes of objects are evident, with the small Galactic globular 
clusters appearing at the low $\sigma_{\infty}$ extreme as circles and triangles giving results for elliptical 
galaxies. The solid line is not a fit to the data but gives the $R_{M}=3 \sigma_{\infty}/a_{0}$ prediction of MONDian 
gravity of Equation (4).

In going from volumetric velocity dispersion profiles to the observable projected velocity dispersion profiles, 
various projection effects intervene, in ways which depend on the details of the degree of central concentration
of the volumetric velocity dispersion profile, and the real space density profiles of the tracers being
analyzed (Hernandez \& Jimenez 2012; Jimenez et al. 2013a; Tortora 2014). Such details of the integration 
implicit in obtaining the line of sight projected profiles
we treat necessarily introduce changing systematics across both classes of systems, from the centrally 
peaked kinematics of globular clusters to the flatter ones of elliptical galaxies. To the above we must add the
systematics inherent
to any comparison across such distinct astrophysical systems over such a range of distances and observational 
techniques as what we are attempting here. We thus find it encouraging of the physical interpretation presented that 
the overall trend evident in Figure 5 can be clearly well represented by the predictions of Equation (4), under the
natural identification of $R_{\sigma}=R_{M}$, the transition to the flat velocity dispersion regime being given by
the length scale of MONDian gravity.

We also performed a weighted linear fit to the data in Figure 5 to equation:

\begin{equation}
\left( \frac{R_{\sigma}}{pc} \right) = A \left( \frac{\sigma_{\infty}}{km/s}
\right)^{n}.
\end{equation}

\noindent 
Such a linear fit to the full data set of Figure 5, shown by the
dashed line, yields $A=0.73 \pm 0.13$ and $n=1.89 \pm 0.32$, in clear accordance with the
MONDian predictions of Equation (5) for $A=0.81$ and $n=2$.
The corresponding fit to only the elliptical galaxies is largely unconstrained,
but if we consider only the globular clusters we obtain $A_{GC}=1.54 \pm 2.35$ and $n_{GC}=1.45 \pm 1.12$.
This is 	not in conflict with the predictions of Equation 5, although the overall
trend is clearly driven by the positions of the two clouds of points coming from
the two distinct scale limited classes considered.

Under Newtonian gravity, the flattening observed in elliptical velocity
dispersion profiles, where tides are not relevant, would be interpreted as
evidence for a dark matter halo, and the consistency of $R_{\sigma} \propto \sigma^{2}$
of Figure (5), as another curious coincidence down to the details of biasing
and stellar feedback.

\begin{figure}
\hskip -10pt \includegraphics[width=8.8cm,height=6.8cm]{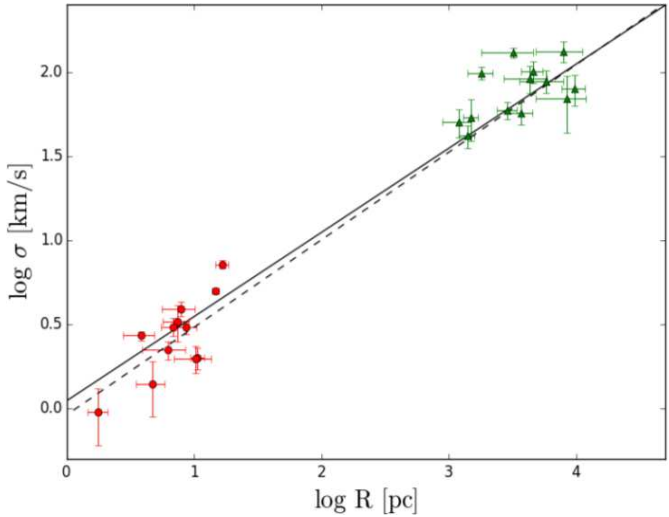}
\caption{$\sigma_{\infty}$ vs. $R_{\sigma}$ values for both of the systems treated, red circles for
  Galactic globular clusters and green triangles for low rotation elliptical galaxies, in a logarithmic plane.
  The solid line is not a fit to the data, it rather shows the MONDian 
expectations of Equation (4) for the predicted scaling of $R_{M}=3 \sigma_{\infty}^{2}/a_{0}$, $R_{M}/pc =
0.81 (\sigma_{\infty}/kms^{-1})^{2}$. A linear regression fit (dashed line) gives $R_{\sigma}=(0.73 \pm 0.13)
\sigma_{\infty}^{1.89 \pm 0.32}$, clearly compatible with the theoretical expectation under the identification of
$R_{\sigma}=R_{M}$. }
\end{figure}

Within MOND as such, for systems under the influence of an external
gravitational field, a regime change is expected according to whether the
internal acceleration of the system in question is smaller or larger than
the external one. In fact, if the external acceleration dominates and is
larger than $a_{0}$, the system being treated will behave in an essentially
Newtonian manner, even if internal accelerations are well below $a_{0}$
(Famaey \& McGaugh 2012). While not relevant for elliptical galaxies,
the collection of Galactic globular clusters
treated here span the above MOND transition, some of the closer ones would
be expected to be affected by this so called external field effect, while
the more distant, e.g. NGC 2419, would be exempt from it, Sanders (2012).
From Figure 5 it is clear that all clusters are consistent with the first
order MONDian expectations of Equation (4), regardless of their
galactocentric distances. Our results hence appear to support generic
MONDian schemes where the external field effect might not appear, or result
in milder distortions than what appear in MOND as such, e.g. Verinde (2016)
or Barrientos \& Mendoza (2016).

It appears that a unique gravitational physics applies, with a clear transition
radius at a $R_{M}$ scale where the decline of the inner Newtonian $R<R_{M}$ region gives way to the
``Tully-Fisher'' $\sigma(R) \propto (G M a_{0})^{1/4}$ of the outer MONDian region, regardless of the fact that
we start with self-gravitating systems of stars at pc scales, to the much larger ellipticals at kpc scales.

Increasing our sample size and extending the sample with other type of pressure supported systems, e.g. dSphs, to
validate our conclusions is evidently a desirable development, which however must be undertaken with care to keep
the number of objects at each class considered relatively balanced, to make sure that the systematics inherent to
any particular class are not biasing conclusions. Similarly, filling the gap present at $\sigma_{\infty}$ values of
between 10 and 40 $km/s$ would be highly desirable.

\section{Conclusions}\label{ccl}

A universal projected velocity dispersion profile of $\sigma(R)=\sigma_{0}e^{-(R/R_{\sigma})^{2}}+\sigma_{\infty}$
has been shown to accurately model two distinct classes of astronomical systems; globular clusters and 
low rotation elliptical galaxies.

MONDian gravity generically predicts $R_{M}=3 \sigma_{\infty}^{2}/a_{0}$, $R_{M}/pc=0.81 (\sigma_{\infty}/km s^{-1})^{2}$.
Data across 7 orders of magnitude in mass give $R_{\sigma}/pc=A(\sigma_{\infty}/km s^{-1})^{n}$, 
with $A=0.73 \pm 0.13$ and $n=1.89 \pm 0.32$.

Thus, in spite of significant observational uncertainties and varying projection effects across apparently very 
different classes of astronomical systems, the identification of the local physical $R_{M}$ radius with the empirical
projected kinematic $R_{\sigma}$ parameter yields results consistent with generic modified gravity expectations, 
where gravitational anomalies in the $a<a_{0}$ regime stem from changes in the physics and not a hypothetical dark matter
component.

\section*{acknowledgements}

The authors acknowledge constructive criticism from an anonymous referee
as important towards reaching a clearer and more complete revised version
of this manuscript.
Xavier Hernandez acknowledges financial assistance from UNAM DGAPA grant IN100814. Reginaldo Durazo acknowledges
financial assistance from a CONACyT scholarship. Bernardo Cervantes Sodi acknowledge financial support through PAPIIT
project IA103517 from DGAPA-UNAM. SFS thanks the CONACYT-125180, DGAPA-IA100815 and DGAPA-IA101217
projects for providing him support in this study.


\begin{thebibliography}{}

\bibitem[Abell(1958)]{1958ApJS....3..211A} Abell, G.~O.\ 1958, \apjs, 3, 211 

\bibitem[Allen et al.(2006)]{2006ApJ...652.1150A} Allen, C., Moreno, E., \& Pichardo, B.\ 2006, \apj, 652, 1150 

\bibitem[Allen et al.(2008)]{2008ApJ...674..237A} Allen, C., Moreno, E., \& Pichardo, B.\ 2008, \apj, 674, 237-246 

\bibitem[Barkhouse et al.(2007)]{2007ApJ...671.1471B} Barkhouse, W.~A., Yee, H.~K.~C., \& L{\'o}pez-Cruz, O.\ 2007, \apj, 671, 1471 

\bibitem[Barrientos \& Mendoza(2016)]{2016arXiv161207970B} Barrientos, E., \& Mendoza, S.\ 2016, arXiv:1612.07970 

\bibitem[Baumgardt(2017)]{2017MNRAS.464.2174B} Baumgardt, H.\ 2017, \mnras, 464, 2174 

\bibitem[Bekenstein(2004)]{2004PhRvD..70h3509B} Bekenstein, J.~D.\ 2004, \prd, 70, 083509 

\bibitem[Binney \& Tremaine(1987)]{1987gady.book.....B} Binney, J., \& Tremaine, S.\ 1987, Princeton, NJ, Princeton University Press, 1987, 747 p.,  

\bibitem[Bower et al.(1992)]{1992MNRAS.254..601B} Bower, R.~G., Lucey, J.~R., \& Ellis, R.~S.\ 1992, \mnras, 254, 601 

\bibitem[Capozziello \& de Laurentis(2011)]{2011PhR...509..167C} Capozziello, S., \& de Laurentis, M.\ 2011, \physrep, 509, 167 

\bibitem[Cappellari et al.(2006)]{2006MNRAS.366.1126C} Cappellari, M., Bacon, R., Bureau, M., et al.\ 2006, \mnras, 366, 1126 

\bibitem[Chae \& Gong(2015)]{2015MNRAS.451.1719C} Chae, K.-H., \& Gong, I.-T.\ 2015, \mnras, 451, 1719 

\bibitem[Dabringhausen et al.(2016)]{Dabinghausen16} Dabringhausen, J., Kroupa, P., Famaey, B., \& Fellhauer, M.\ 2016, \mnras, 463, 1865 

\bibitem[Desmond(2017)]{2017MNRAS.464.4160D} Desmond, H.\ 2017, \mnras, 464, 4160 


\bibitem[Emsellem et al.(2011)]{2011MNRAS.414..888E} Emsellem, E., Cappellari, M., Krajnovi{\'c}, D., et al.\ 2011, \mnras, 414, 888 


\bibitem[Emsellem et al.(2004)]{2004MNRAS.352..721E} Emsellem, E., Cappellari, M., Peletier, R.~F., et al.\ 2004, \mnras, 352, 721 


\bibitem[Famaey \& McGaugh(2012)]{2012LRR....15...10F} Famaey, B., \& McGaugh, S.~S.\ 2012, Living Reviews in Relativity, 15, 10 


\bibitem[Famaey \& Binney(2005)]{2005MNRAS.363..603F} Famaey, B., \& Binney, J.\ 2005, \mnras, 363, 603 


\bibitem[Gentile et al.(2010)]{2010A&A...509A..97G} Gentile, G., Famaey, B., Angus, G., \& Kroupa, P.\ 2010, \aap, 509, A97 


\bibitem[Hernandez \& Jim{\'e}nez(2012)]{2012ApJ...750....9H} Hernandez, X., \& Jim{\'e}nez, M.~A.\ 2012, \apj, 750, 9 


\bibitem[Hernandez et al.(2013)]{2013ApJ...770...83H} Hernandez, X., Jim{\'e}nez, M.~A., \& Allen, C.\ 2013a, \apj, 770, 83 


\bibitem[Hernandez et al.(2013)]{2013MNRAS.428.3196H} Hernandez, X., Jim{\'e}nez, M.~A., \& Allen, C.\ 2013b, \mnras, 428, 3196 


\bibitem[Jim{\'e}nez et al.(2013)]{2013ApJ...768..142J} Jim{\'e}nez, M.~A., Garcia, G., Hernandez, X., \& Nasser, L.\ 2013, \apj, 768, 142 


\bibitem[Jones \& Forman(1999)]{1999ApJ...511...65J} Jones, C., \& Forman, W.\ 1999, \apj, 511, 65 


\bibitem[Kennedy(2014)]{2014MNRAS.445.4446K} Kennedy, G.~F.\ 2014, \mnras, 445, 4446 


\bibitem[K{\"u}pper et al.(2010)]{2010MNRAS.401..105K} K{\"u}pper, A.~H.~W., Kroupa, P., Baumgardt, H., \& Heggie, D.~C.\ 2010, \mnras, 401, 105


\bibitem[Lane et al.(2011)]{2011A&A...530A..31L} Lane, R.~R., Kiss, L.~L., Lewis, G.~F., et al.\ 2011, \aap, 530, A31 


\bibitem[Lane et al.(2010)]{2010MNRAS.401.2521L} Lane, R.~R., Kiss, L.~L., Lewis, G.~F., et al.\ 2010, \mnras, 401, 2521 


\bibitem[Lane et al.(2009)]{2009MNRAS.400..917L} Lane, R.~R., Kiss, L.~L., Lewis, G.~F., et al.\ 2009, \mnras, 400, 917 


\bibitem[Lane et al.(2010)]{2010MNRAS.406.2732L} Lane, R.~R., Kiss, L.~L., Lewis, G.~F., et al.\ 2010, \mnras, 406, 2732 

\bibitem[Lelli et al.(2016)]{2016arXiv161008981L} Lelli, F., McGaugh, S.~S., Schombert, J.~M., \& Pawlowski, M.~S.\ 2016, accepted for publication in ApJ, arXiv:1610.08981 

\bibitem[Massari et al.(2016)]{2016arXiv161200183M} Massari, D., Posti, L., Helmi, A., Fiorentino, G., \& Tolstoy, E.\ 2016, accepted for publication in A\&A Letters, arXiv:1612.00183 
  
\bibitem[McGaugh(2012)]{2012AJ....143...40M} McGaugh, S.~S.\ 2012, \aj, 143, 40 

\bibitem[McGaugh et al.(2016)]{2016PhRvL.117t1101M} McGaugh, S.~S., Lelli, F., \& Schombert, J.~M.\ 2016, Physical Review Letters, 117, 201101 


\bibitem[McLaughlin \& van der Marel(2005)]{2005ApJS..161..304M} McLaughlin, D.~E., \& van der Marel, R.~P.\ 2005, \apjs, 161, 304 

  
\bibitem[Mendoza et al.(2013)]{2013MNRAS.433.1802M} Mendoza, S., Bernal, T., Hernandez, X., Hidalgo, J.~C., \& Torres, L.~A.\ 2013, \mnras, 433, 1802 


\bibitem[Mendoza et al.(2011)]{2011MNRAS.411..226M} Mendoza, S., Hernandez, X., Hidalgo, J.~C., \& Bernal, T.\ 2011, \mnras, 411, 226 


\bibitem[Milgrom(1983)]{Milgrom83} Milgrom, M.\ 1983, \apj, 270, 365 


\bibitem[Milgrom(1984)]{1984ApJ...287..571M} Milgrom, M.\ 1984, \apj, 287, 571 


\bibitem[Richtler et al.(2008)]{2008A&A...478L..23R}Richtler, T., Schuberth, Y., Hilker, M., et al.\ 2008, \aap, 478, L23 


\bibitem[Rodrigues et al.(2014)]{2014MNRAS.445.3823R} Rodrigues, D.~C., de Oliveira, P.~L., Fabris, J.~C., \& Gentile, G.\ 2014, \mnras, 445, 3823 


\bibitem[Samurovi{\'c}(2014)]{2014A&A...570A.132S} Samurovi{\'c}, S.\ 2014, \aap, 570, A132 


\bibitem[Samurovi{\'c}(2016)]{2016Ap&SS.361..199S} Samurovi{\'c}, S.\ 2016, \apss, 361, 199 


\bibitem[S{\'a}nchez et al.(2016)]{2016A&A...594A..36S} S{\'a}nchez, S.~F., Garc{\'{\i}}a-Benito, R., Zibetti, S., et al.\ 2016a, \aap, 594, A36 


\bibitem[S{\'a}nchez et al.(2016)]{2016RMxAA..52..171S} S{\'a}nchez, S.~F., P{\'e}rez, E., S{\'a}nchez-Bl{\'a}zquez, P., et al.\ 2016b, \rmxaa, 52, 171 


\bibitem[S{\'a}nchez et al.(2012)]{2012A&A...538A...8S} S{\'a}nchez, S.~F., Kennicutt, R.~C., Gil de Paz, A., et al.\ 2012, \aap, 538, A8 


\bibitem[Sanders(2012)]{2012MNRAS.419L...6S} Sanders, R.~H.\ 2012, \mnras, 419, L6 


\bibitem[Sanders \& McGaugh(2002)]{2002ARA&A..40..263S} Sanders, R.~H., \& McGaugh, S.~S.\ 2002, \araa, 40, 263 


\bibitem[Scarpa \& Falomo(2010)]{2010A&A...523A..43S} Scarpa, R., \& Falomo, R.\ 2010, \aap, 523, A43 


\bibitem[Scarpa et al.(2011)]{2011A&A...525A.148S} Scarpa, R., Marconi, G., Carraro, G., Falomo, R., \& Villanova, S.\ 2011, \aap, 525, A148 


\bibitem[Scarpa et al.(2007)]{2007A&A...462L...9S} Scarpa, R., Marconi, G., Gilmozzi, R., \& Carraro, G.\ 2007, \aap, 462, L9 


\bibitem[Scarpa et al.(2007)]{2007Msngr.128...41S} Scarpa, R., Marconi, G., Gilmozzi, R., \& Carraro, G.\ 2007, The Messenger, 128,  


\bibitem[Serra et al.(2011)]{2011MNRAS.412..800S} Serra, A.~L., Diaferio, A., Murante, G., \& Borgani, S.\ 2011, \mnras, 412, 800 


\bibitem[Swaters et al.(2010)]{2010ApJ...718..380S} Swaters, R.~A., Sanders, R.~H., \& McGaugh, S.~S.\ 2010, \apj, 718, 380 


\bibitem[Tian \& Ko(2016)]{2016MNRAS.462.1092T} Tian, Y., \& Ko, C.-M.\ 2016, \mnras, 462, 1092 


\bibitem[Tortora et al.(2014)]{2014MNRAS.438L..46T} Tortora, C., Romanowsky, A.~J., Cardone, V.~F., Napolitano, N.~R., \& Jetzer, P.\ 2014, \mnras, 438, L46 


\bibitem[Varela et al.(2004)]{2004A&A...420..873V} Varela, J., Moles, M., M{\'a}rquez, I., et al.\ 2004, \aap, 420, 873 


\bibitem[Verlinde(2016)]{2016arXiv161102269V} Verlinde, E.~P.\ 2016, arXiv:1611.02269 


\bibitem[Walcher et al.(2014)]{2014A&A...569A...1W} Walcher, C.~J., Wisotzki, L., Bekerait{\'e}, S., et al.\ 2014, \aap, 569, A1 


\bibitem[Wolf et al.(2010)]{2010MNRAS.406.1220W} Wolf, J., Martinez, G.~D., Bullock, J.~S., et al.\ 2010, \mnras, 406, 1220 


\bibitem[Yahil \& Vidal(1977)]{1977ApJ...214..347Y} Yahil, A., \& Vidal, N.~V.\ 1977, \apj, 214, 347 


\bibitem[York et al.(2000)]{2000AJ....120.1579Y} York, D.~G., Adelman, J., Anderson, J.~E., Jr., et al.\ 2000, \aj, 120, 1579 




\end{thebibliography}
\end{document}